\documentclass[useAMS,usenatbib]{mn2e}
\usepackage{amsmath}%
\usepackage{amsfonts}%
\usepackage{amssymb}
\usepackage{graphicx, natbib, wasysym, subfiles, aas_macros}

\newcommand{\msun}{\ensuremath{{\rm M}_{\astrosun}}}
\newcommand{\mearth}{\ensuremath{{\rm M}_{\oplus}}}

\newcommand{\mdot}{\ensuremath{\dot{\rm M}}}

\newcommand{\tr}{\ensuremath{r}}
\newcommand{\tz}{\ensuremath{z}}
\newcommand{\trho}{\ensuremath{\rho}}
\newcommand{\tp}{\ensuremath{p}}
\newcommand{\tT}{\ensuremath{T}}
\newcommand{\ttime}{\ensuremath{t}}
\newcommand{\tv}{\ensuremath{v}}

\newcommand{\trcav}{\ensuremath{\tr_{\rm cav}}}
\newcommand{\AU}{AU}

\newcommand{\eqn}[1]{Eq. \ref{#1}}
\newcommand{\fig}[1]{Fig. \ref{#1}}

\newcommand{\parti}[2]{\partial_{#2} #1}
\renewcommand{\vec}[1]{\ensuremath{\boldsymbol{#1}}} 
\DeclareMathAlphabet{\mathbfsf}{\encodingdefault}{\sfdefault}{bx}{sl}
\newcommand{\tensor}[1]{\ensuremath{\mathbfsf{#1}}} 

\title[Planet trapping at disc-cavity boundaries]{3D Simulations of Planet Trapping at Disc-Cavity Boundaries}

\author[Romanova et al.]{\parbox{\textwidth}{M. M.~Romanova$^{1,2}$\thanks{E-mail of
corresponding author: \texttt{romanova@astro.cornell.edu}},  P. S.
~Lii$^{2}$, A. V.~Koldoba$^{3}$, G. V.~Ustyugova$^{4}$,
A. A. Blinova$^{1,2}$,  R. V. E.~Lovelace$^{1,2,5}$, L.
Kaltenegger$^{1,2}$
}\vspace{0.4cm}\\
\parbox{\textwidth}{ 
$^{1}$Department of Astronomy, Cornell University, Ithaca, NY 14853-6801\\
$^{2}$Carl Sagan Institute, Cornell University, Ithaca, NY 14853-6801\\
$^{3}$Moscow Institute of Physics and Technology, Dolgoprudny, Moscow Region, 141700, Russia \\
$^{4}$Keldysh Institute for Applied Mathematics, Moscow, 125047,
Russia \\
$^{5}$Department of Applied and Engineering Physics, Cornell
University, Ithaca, NY 14853-6801}}
\date{\today}

\pagerange{\pageref{firstpage}--\pageref{lastpage}} \pubyear{2002}

\begin{document}
\label{firstpage}

\maketitle

\begin{abstract}

Inward migration of low-mass planets and embryos of giant planets
can be stopped at the disc-cavity boundaries due to  co-orbital
corotation torque. We performed the first global three-dimensional
(3D) simulations of planet migration at the disc-cavity boundary,
and have shown that the boundary is a robust trap for low-mass
planets and embryos. A protoplanetary disc may have several such
trapping regions at various distances from the star, such as at
the edge of the stellar magnetosphere, the inner edge of the dead
zone, the dust-sublimation radius and the snow lines. Corotation
traps located at different distances from a star, and moving
outward during the disc dispersal phase, may possibly explain the
observed homogeneous distribution of low-mass planets with
distance from their host stars.



\end{abstract}

\begin{keywords}
accretion discs, hydrodynamics, planet-disc interactions,
protoplanetary discs
\end{keywords}

\section{Introduction}

Thousands of planets were discovered by \textit{COROT, Kepler,
K2}, and other space and ground-based telescopes (3,989 planets,
 February 19 2019, see   exoplanet.eu; see also, e.g.
\citealt{BoruckiEtAl2010}).
The current distribution of planets reflects the complex history
of their formation in protoplanetary discs, and their dynamical
interaction with their discs, host stars, and other planets (e.g.
\citealt{WynnFabrycky2015}).

Most of the planets observed at the distances of 0.1-1 AU from
their host stars have relatively low masses, $M_p\sim
1-30M_\oplus$. These planets are almost homogeneously distributed
\footnote{After removing the  selection effect
\citep{FressinEtAl2013}.} with respect to their orbital periods
(e.g. \citealt{FressinEtAl2013,WynnFabrycky2015}). These planets
have different radii in the range of $1 R_\oplus\lesssim
R_p\lesssim 4 R_\oplus$ (see Fig. 1 from review by
\citealt{Kaltenegger2017}; see also \citealt{ZengEtAl2016}).
Some of them
 have higher densities and are expected to be predominantly rocky,
while others have lower densities and are expected to be
predominantly gaseous. Gaseous planets are expected to form
when a significant amount of gas is present in the disc during
planet formation. If a disc is present during planet formation,
then the newly-formed planets interact with the disc
gravitationally and migrate. Their migration time scale is
expected to be lower than the time scale of planet formation (e.g.
\citealt{LinPapaloizou1986}).

In addition, a significant number of more massive, Jupiter-sized
planets are present in the planetary data set. These planets are
expected to form in gaseous discs,
starting from seed rocky cores (embryos) of a few Earth masses
(e.g. \citealt{PollackEtAl1996}). These rocky cores can also
migrate rapidly and fall onto the star before  a planet can form.

Low-mass planets and rocky embryos of giant planets are expected
to migrate in the Type I regime, where the planet does not open a
gap. The torque acting on a planet consists of two components: the
differential Lindblad torque and the corotation torque
\citep{GoldreichTremaine1979,GoldreichTremaine1980,LinPapaloizou1979,Ward1986,
Ward1997}. The Lindblad torque arises from the tidal interaction
of a planet with the disc, which leads to the formation of two
spiral density waves that exert positive and negative torques on
the planet.  The cumulative Lindblad torque is typically negative.
The corotation torque arises from the interaction of a planet with
the co-orbital matter; it is usually positive. In a typical
accretion disc, where the surface density smoothly decreases with
distance from the star (e.g. \citealt{Hayashi1981}), the negative
Lindblad torque is usually larger (in absolute value) than the
corotation torque, and the planet migrates inward towards the
star. For a planet located at a few AU, the time scale of
migration is smaller than both the time scale of planet formation
and the life time of the disc. Therefore, many planets and rocky
embryos are expected to migrate to small radii, and  may fall onto
their host stars (e.g. \citealt{LinPapaloizou1986}, see also
review paper by \citealt{KleyNelson2012}). This picture of
migration does not correspond to the observed dispersed
distribution of low-mass planets with distance from their host
stars (e.g. \citealt{WynnFabrycky2015}).

In recent years,
different factors   that may slow down or reverse migration were
taken into account, such as (a) density and temperature gradients
in the disc (e.g.
\citealt{Masset2001,DAngeloLubow2010,LegaEtAl2014,MassetBenitez2016}),
(b) turbulence in the disc (e.g. \citealt{Nelson2005}), (c)
magnetic fields (e.g.
\citealt{Terquem2003,FromangEtAl2005,BaruteauEtAl2011,CominsEtAl2016,
McNallyEtAl2017,McNallyEtAl2018}),  and other factors (see review
papers by
\citealt{Masset2008,KleyNelson2012,BaruteauEtAl2014,BaruteauEtAl2016}).

In particular, it has been shown that, if some  parts of the disc
have positive density gradients, then the positive corotation
torque dominates over the negative Lindblad torque, and the
planets can either migrate outward or the migration may stall
(e.g.
\citealt{Masset2001,TanakaEtAl2002,MassetEtAl2006a,Masset2008,PaardekooperPapaloizou2009a}).

Strong positive density gradients are expected at the transition
region between the disc and the inner gap or cavity, where the
density is lower than that in the disc \citep{MassetEtAl2006a}.
Such cavities can be associated with transition regions in the
disc, such as the magnetospheric boundary, a dead zone, the dust
sublimation radius, and the radius of the snow line (e.g.
\citealt{MassetEtAl2006a,MorbidelliEtAl2008,IdaLin2008,
HasegawaPudritz2011}).

\begin{figure*}
     \includegraphics[width=9.0cm]{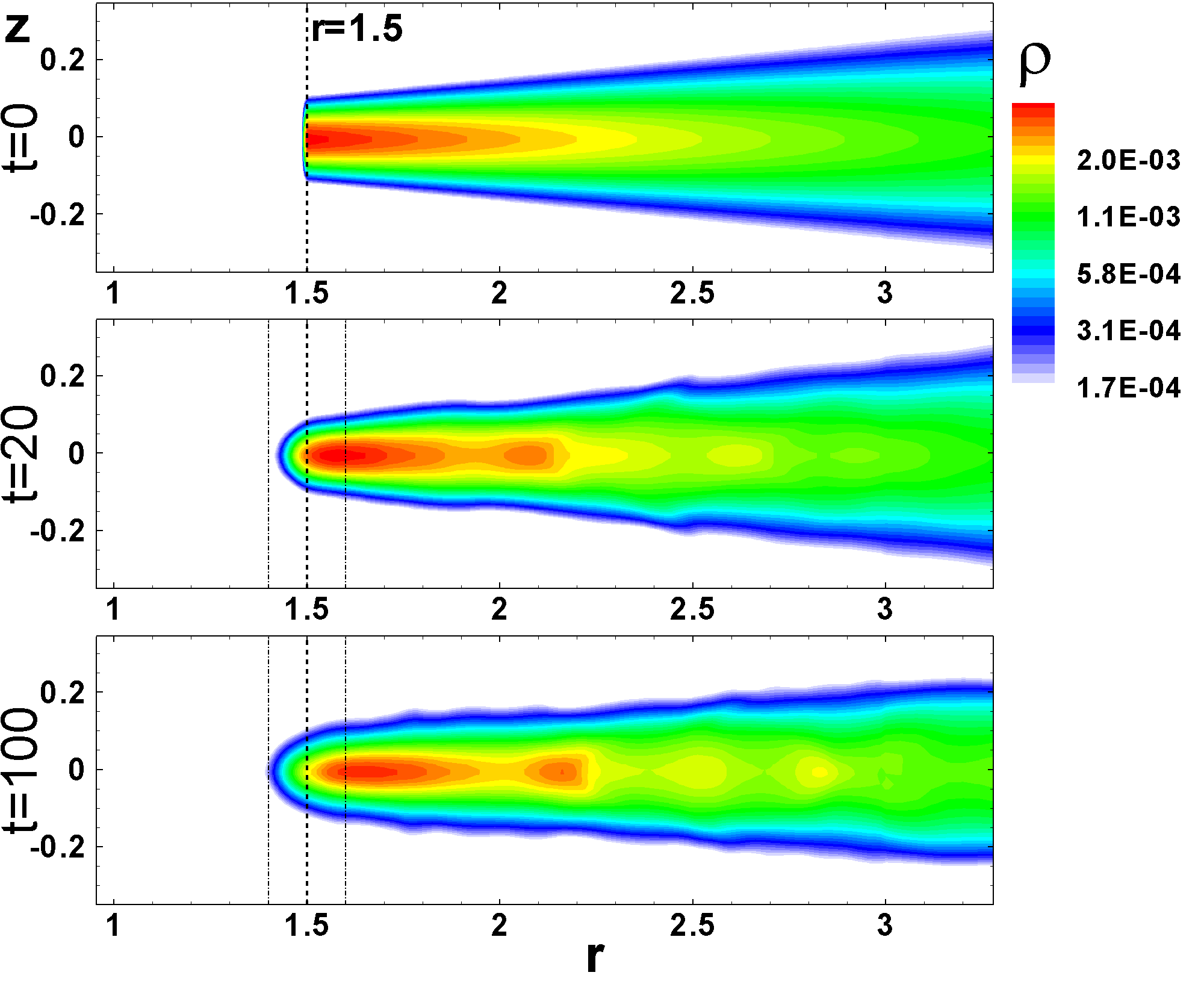} 
     \includegraphics[width=8.0cm]{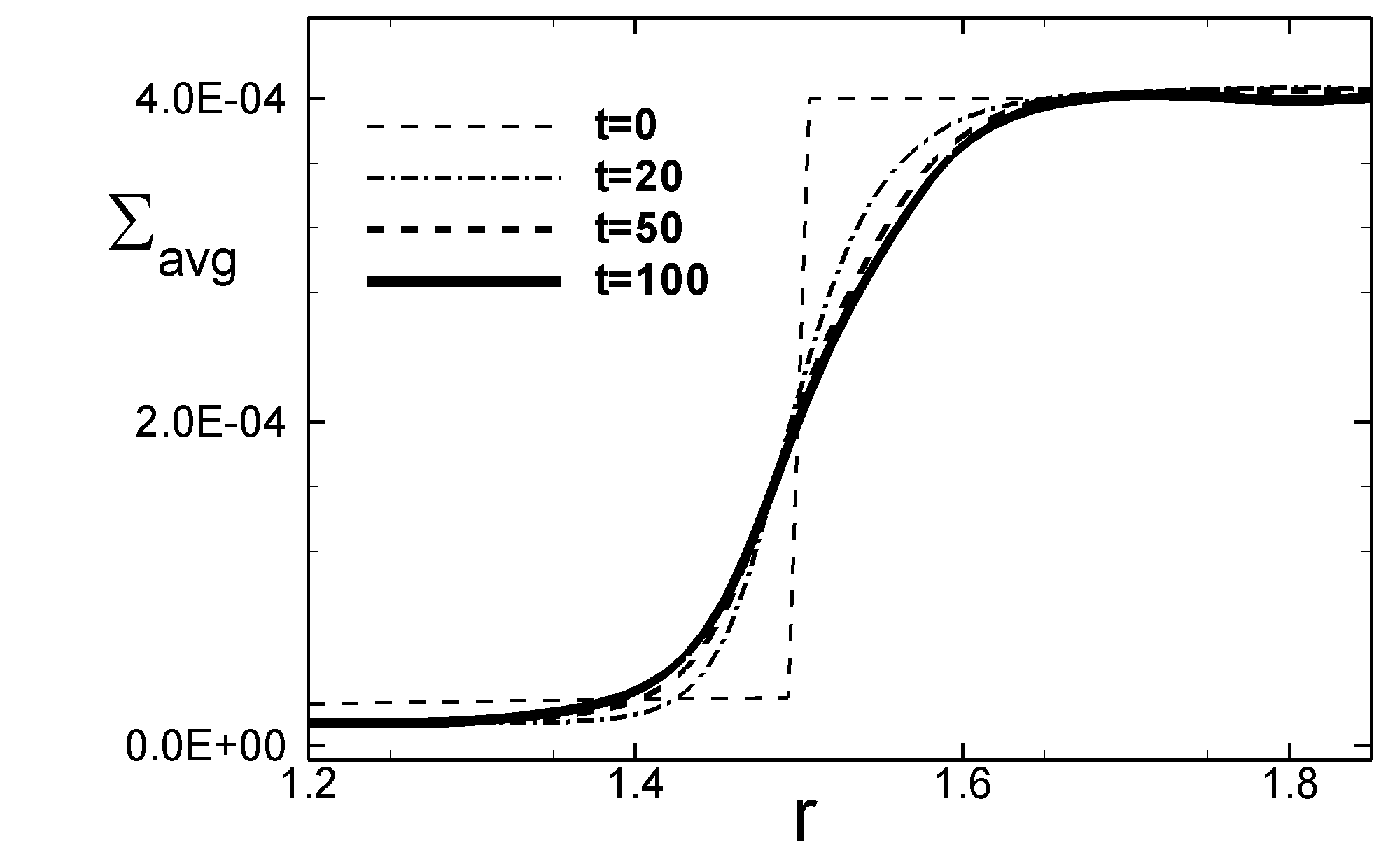} 
     \caption{\textit{Left panels:} $rz$-slices through the disc showing the density distribution (colour background)
  at $t=0, 20, 100$.
     Vertical line in the top panel shows the position of the sharp density gradient
     at $t=0$. Dashed lines in the middle and bottom panels show approximate boundaries
     for the large positive gradient ($1.4\lesssim r \lesssim 1.6$) at $t=20$ and $t=100$.
     \textit{Right panel:} radial distribution of the azimuthally-averaged surface density,
     $\Sigma_{\rm avg}$, at the disc-cavity boundary at $t=0, 20, 50, 100$.
     \label{fig:xz}}
\end{figure*}

\citet{MassetEtAl2006a} studied the effects of corotation torque
on the migration of planets in the vicinity of a cavity edge using
2D hydrodynamic simulations. In their study, low-mass planets
(1.3-15 $M_\oplus$) were allowed to migrate in an accretion disc
with a lower-density cavity. Far from the cavity boundary,
planets migrated inward as usual due to the differential Lindblad
torque. However, as the planets approached the density transition,
the authors found that the corotation torque contributed a
positive torque, creating a stable \textit{planet trap} where the
net torque is zero and the inward migration is halted. Instead of
migrating into the cavity (as would be expected if the Lindblad
resonance were dominant),
the migration of planets halts while they are still in the disc
due to the action of the corotation torque. The authors have shown
that planets are stably trapped at the boundary, and  when the
boundary changes its position the trapped planets move together
with the boundary.

\citet{MorbidelliEtAl2008} investigated the planet trap mechanism
as a way of avoiding the inspiral problem, focusing on the case
where the planet embryos are trapped at the cavity boundary,
allowing for a gradual buildup of giant-planet cores.

\citet{LiuEtAl2017} studied the trapping of low-mass planets at a
sharp disc-magnetosphere boundary using a semi-analytical
approach. Here, the authors have shown that the one-sided positive
corotation torque at the boundary is larger than the one-sided
negative Lindblad torque for a wide range of low-mass planets. The
authors considered evolving discs, where the disc-magnetosphere
boundary moves outward during the disc dispersal phase. They
suggested that the planets move with the boundary as long as the
surface density in the disc is sufficiently high. The authors
proposed this mechanism for explaining the observed dispersed
distribution of planets at $r\lesssim 1$ AU
\footnote{\citet{LiuEtAl2017} have also shown that, in the case of
two planets migrating simultaneously in the disc, the outward
motion of the inner (trapped) planet takes the system out of the
mean-motion resonance, which is expected in the case of migrating
planets but is usually not observed (e.g.
\citealt{CresswellNelson2006}).}.

In this study, we investigate for the first time the migration of
low-mass planets and embryos at the disc-cavity boundary  using
global three-dimensional (3D) simulations. In Sec.
\ref{sec:model}, we discuss the problem setup. In Sec.
\ref{sec:numerical model}, we show the details of our numerical
model. In Sec. \ref{sec:migration-boundary}, we describe our
results. In Sec. \ref{sec:applic}, we discuss the applications of
our model. In Sec. \ref{sec:conclusion},  we conclude. In
Appendixes \ref{sec:appen-aspect ratio} and
\ref{sec:appen-viscosity}, we show the results of our simulations
for discs with different aspect ratios and at different
viscosities in the disc, respectively.

\section{Problem setup}
\label{sec:model}

We consider  the migration of the low-mass planets
at the disc-cavity
boundary using global 3D simulations.

\paragraph*{Initial conditions.}

The simulation region consists of a disc, which is truncated at
the radius of $r=r_d=r_{\rm cav}$,
 a low-density cavity located at $r<r_{\rm cav}$,
and a low-density corona above and below the disc.
 Initially, the matter in the
disc is dense and cold, while the cavity and corona are
 filled with hot, lower density gas. In most of the simulation
runs, we take a disc with an aspect ratio of $h=H/r=0.03$, which
is determined at time $t=0$ at the inner edge of the disc,
$r=r_{\rm cav}$. The scale height of the disc is $H=(c_s/v_{\rm
K}) r$, where $c_s$ and $v_K$ are the sound speed and Keplerian
velocity, respectively \footnote{In the test cases, we also study
migration in discs with larger aspect ratios: $h=H/r=0.05$ and
$0.1$ (see Appendix \ref{sec:appen-aspect ratio}).}.

We calculate the equilibrium distribution of density and pressure
in the disc and corona using the following method. First, we
determine the equilibrium in the equatorial plane. The initial
radial density and pressure distributions are given by
\begin{equation}
\trho(\tr, 0) = \left\{
\begin{array}{lr}
\trho_{\rm cav} \left(\frac{\tr}{\trcav}\right)^{-n} & {\rm if}\ \tr < \trcav \\
\trho_{\rm d} \left(\frac{\tr}{\trcav}\right)^{-n} & {\rm if}\ \tr
\ge \trcav
\end{array}
\right. \quad, \label{eqn_rhoinit}
\end{equation}
and \begin{equation} \tp(\tr, 0) = \tp_{\rm d}
\left(\frac{\tr}{\trcav}\right)^{-l} \quad. \label{eqn_pressure}
 \end{equation}
Here, $\trho_{\rm cav}$, $\trho_{\rm d}$ and $\tp_{\rm d}$ set the
cavity density, disc density and pressure, respectively, at $\tr =
\trcav$ (in the equatorial plane). Parameters $n$ and $l$ specify
the radial profile of the disc density and pressure, respectively.
We use the values $n = 1.5$ and $l = 1.5$ in all simulation runs.

Near the inner edge of the disc ($r=r_d$), the pressure in the
inner disc should be equal to the pressure in the cavity, $p_{\rm
d}=p_{\rm cav}$. In the equatorial plane, the dimensionless
temperature is related to density and pressure by the ideal gas
law, ${\cal R}\tT(\tr, 0) = \tp(\tr, 0)/\trho(\tr, 0)$
\footnote{Our initial disc is locally isothermal, that is, the
temperature only depends on radius. Subsequently, at $t>0$, we
calculate the temperature distribution using the energy
equation.}.

Next, we assume that there is a hydrostatic equilibrium in the
vertical direction and build the 3D distribution of density and
pressure:
\begin{equation}
\trho(\tr, \tz) = \trho(r,0)
\exp\left(\frac{\Phi(\tr,0)-\Phi(\tr,\tz)}{{\cal
R}\tT(\tr,0)}\right)~, \label{eqn_potential}
\end{equation}
 where $\Phi(\tr,\tz) = -GM_*/(\tr^2+\tz^2)^{1/2}$ is the
gravitational potential of the star. The expression for pressure
is analogous. The azimuthal velocity $\tv_{\phi}$ is set by the
balance of gravity and pressure gradient forces in the radial
direction:
\begin{equation}
\tv_\phi(\tr,\tz) = \sqrt{r \left( \frac{\partial \Phi}{\partial
r} + \frac{1}{\rho} \frac{\partial p}{\partial r} \right)}~.
\label{eq:vphi}
\end{equation}
These formulae allow us to start from a quasi-equilibrium
configuration for the disc and the cavity.

We calculate the surface density distribution in the disc
 by integrating the volume density $\rho$ in the $z-$direction:
\begin{equation}
\Sigma=\int{\rho dz}\propto H \rho \propto \frac{c_s \rho}{\Omega}
\propto \frac{\sqrt{p \rho}}{\Omega} \propto r^{\frac{3-n-l}{2}} .
\label{eqn:_Sigma}
\end{equation}
Here, $c_s\propto\sqrt{p/\rho}$ is the sound speed. The slopes of
$n = 1.5$ and $l = 1.5$ in the equatorial density and pressure
distributions result in a flat surface density profile,
$\Sigma=const$ the disc. Note that the corotation torque strongly
depends on the local positive density gradient at the disc-cavity
boundary, while the density distribution at larger (and smaller)
distances from the cavity is not as important. This is why we
choose this initial density distribution and fixed it in all
simulation runs.

\paragraph*{Dimensionalization.}
The simulations are dimensionless and can be applied to cavities
located at different distances from the star. For
dimensionalization, we choose a reference scale,  $r_0$, and
place the cavity at the radius of $r_{\rm cav}=1.5 r_0$.
The reference mass is set to the mass of the star, $M_0 = M_*$.
The reference velocity is given by $v_0 = \sqrt{GM_*/r_0}$,
corresponding to the Keplerian orbital velocity at $r=r_0$. We
measure time in Keplerian periods of rotation at $r=r_0$: $P_0 = 2
\pi r_0/v_0$. The reference density is $\rho_0 = M_0 / r_0^3$ and
the reference surface density is $\Sigma_0 = M_0 / r_0^2$. The
reference pressure is $p_0 = \rho_0 v_0^2$ and the reference
temperature is $T_0 = p_0/(\mathcal{R} \rho_0)$, where
$\mathcal{R}$ is the specific gas constant.

The mass of the planet is given by $M_p = q_p M_0$. In this paper,
we study the migration of planets with masses $M_p = 5 \mearth$
and $M_p = 15 \mearth$, which correspond to dimensionless masses
$q_p = 1.5\times 10^{-5}$ and $q_p = 4.5\times 10^{-5}$,
respectively.

For each dimensionless quantity $\widetilde{Q}$, we recover the
physical value $Q$ by multiplying $\widetilde{Q}$ by the
corresponding reference value, $Q_0$: $Q = \widetilde{Q} Q_0$. For
example, in a disc around a 1 \msun\ star with a low-density
cavity extending out to $r_{\rm cav}=1$ AU, the reference mass is
$M_0 = 1 \msun$, the reference length is $r_0 = 0.66$ AU, the
reference velocity is $36.5$ km s$^{-1}$, the reference period is
$P_0 = 199$ d, and the reference surface density is $\Sigma_0 = 2
\times 10^7$ g cm$^{-2}$. To obtain the physical values of these
quantities, their reference values must then be scaled by the
dimensionless values obtained from the simulations.

Our dimensionless simulations are applicable to cavities located
at different distances from the star. Tab. \ref{tab:units} shows
our reference values and the initial values used in our model. We
give examples of reference values for cavities located at $r_{\rm
cav}=0.1$ AU, $0.5$ AU and $1.0$ AU from the star. Corresponding
reference values for distance are $r_0=r_{\rm cav}/1.5 \approx
0.67, 0.33$ and $0.067$ AU, respectively.

\paragraph*{Initial values in the model.}
The characteristic mass of the disc is $M_{\rm d0} = q_d M_0$,
where $q_d=4\times 10^{-4}$. Initial value of the surface density
in the simulations is $\Sigma_d (t=0)=q_d \Sigma_0$, and therefore
the initial dimensionless value of the surface density is
$\widetilde\Sigma_d=q_d=4\times 10^{-4}$.  The initial
dimensionless values of the surface density in the disc and in the
cavity are: $\widetilde\Sigma_{\rm d}=4\times 10^{-4}$ and
$\widetilde\Sigma_{\rm cav}\approx 0.072 \widetilde\Sigma_{\rm
d}$.


The initial  dimensionless  values of the volume density in the
disc and cavity at the reference point ($r=r_{\rm cav}$) are taken
to be: $\widetilde\trho_{\rm d} = 3.35 \times 10^{-3}$
 and $\widetilde\trho_{\rm cav} = 10^{-2}
\widetilde\rho_{\rm d}=3.35 \times 10^{-5}$ \footnote{Note that
these values are different in models with different values of
thickness of the disc, $h$, see Appendix \ref{sec:appen-aspect
ratio}.}. Initial pressure is $\widetilde \tp_{\rm d} = 2.6 \times
10^{-6}$.

One can obtain dimensional values, using reference values from
Tab. \ref{tab:units} and by multiplying dimensionless values by
reference values $\rho_0$ and $p_0$. For example, for cavity,
located at $r_{\rm cav}=1$ AU, we have $\rho_0=2\times 10^{-6}$ g
cm$^{-3}$ and obtain dimensional values $\rho_{\rm
d}=6.7\times10^{-9}$ g cm$^{-3}$  and $\rho_{\rm
cav}=6.7\times10^{-11}$ g cm$^{-3}$.

Taking into account the reference surface density
$\Sigma_0=2\times 10^7$ g cm$^{-2}$ (at $r_0=0.67$ AU), one
obtains dimensional values $\Sigma_{\rm d}=8\times 10^3$ g
cm$^{-2}$ and $\Sigma_{\rm cav}=5.6\times 10^2$ g cm$^{-2}$. The
bottom three rows of Tab. \ref{tab:units} show the initial values
of the disc mass, density and surface density in the disc for
cavities located at different distances from the star. The right
column shows different values obtained for the Minimum-Mass Solar
Nebula (MMSN) at the distance of $0.67$ AU. As an example, we used
the \citet{Hayashi1981} MMSN, where the surface density
distribution $\Sigma\approx 1700 (r/1\rm{AU})^{-3/2}$ g cm$^{-2}$.

In this study, all of the values are given in terms of
dimensionless units, except where explicitly assigned physical
units. Subsequently in the text, we do not place tilde's above
dimensionless values.

\begin{table*}
\begin{tabular}[]{ cc | ccc | c }
\hline \hline \multicolumn{2}{c}{Reference Unit} &
\multicolumn{3}{c}{Reference Values}    & MMSN  \\ \hline
Reference mass of the star       & $M_0$ [$\msun$]      & 1                & 1                 & 1                & 1        \\
Position of the disc-cavity boundary & $r_{\rm cav}$ [AU] & 0.1 &
0.5 & 1 & 1\\
Reference distance  $r_0=r_{\rm cav}/1.5$           & $r_0$ [AU]            & 0.067            & 0.33              & 0.67             & 0.67  \\
Reference velocity                    & $v_0$ [km s$^{-1}$]          & 115              & 51.6              & 36.5             & 36.5     \\
Reference period                      & $P_0$ [days]          & 6.29             & 70.3              & 199              & 199      \\
Reference density           & $\rho_0$ [g cm$^{-3}$]   & $2.0 \times 10^{-3}$  & $1.6 \times 10^{-5}$   & $2.0 \times 10^{-6}$ \\
Reference surface density   & $\Sigma_0$ [g cm$^{-2}$] & $2.0 \times 10^9$     & $8.0 \times 10^7$      & $2.0 \times 10^7$  &          \\
Reference pressure          & $p_0$ [g cm$^{-1}$ s$^{-2}$]    & $2.6 \times 10^{11}$  & $4.3 \times 10^8$      & $2.7 \times 10^7$  &          \\
\hline  &                &                  &    Initial values in the model               &                   &    MMSN        \\
\hline
Reference mass of the disc  $M_{\rm d0}=4\times10^{-4} M_0$  & $M_{\rm d0}$ [$\msun$]    & $4.0\times 10^{-4}$   & $4.0\times 10^{-4}$    & $4.0\times 10^{-4}$    &          \\
Initial density in the disc at $r=r_{\rm cav}$  & $\rho_d$     [g cm$^{-3}$]   & $8.0 \times 10^{-7}$  & $6.4 \times 10^{-9}$   & $8.0 \times 10^{-10}$  & \\
Initial surface density in the disc     & $\Sigma_d$   [g cm$^{-2}$]   & $8.0 \times 10^5$     & $3.2 \times 10^4$      & $8.0 \times 10^3$  & $1.7 \times 10^3$\\
\hline \hline
\end{tabular}
\caption{Top rows: reference values calculated for different
positions of the disc-cavity boundary $r_{\rm cav}$. Bottom rows:
the reference mass of the disc, $M_{d0}$ and initial values of
density and surface density taken in the model.   The last column
shows the typical values from the model of the MMSN at the
distance of $1$ AU \citep{Hayashi1981}. \label{tab:units}}
\end{table*}

\begin{figure}
    \centering
        \includegraphics[width=0.45\textwidth]{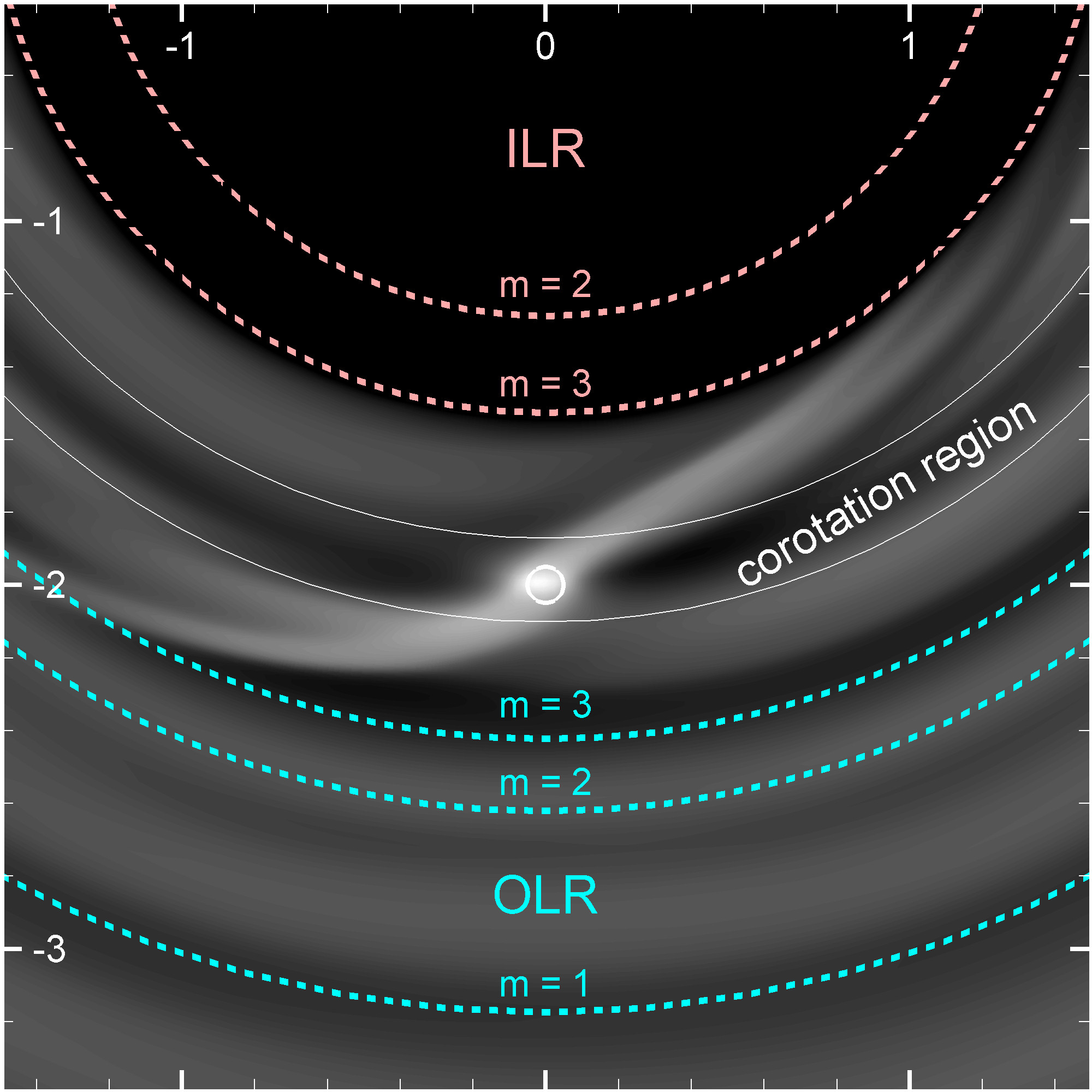}
    \caption{A schematic diagram of various resonance locations for a planet of mass $M_p=15 M_\oplus$.
    The plot shows the corotation region (plotted here as $r_p \pm 3 r_{\rm Hill}$) and the locations of the $m = 1-3$ Lindblad resonances.}
    \label{fig:schematic}
\end{figure}

\section{Numerical model}
\label{sec:numerical model}

 Our code has a hydrodynamics module that models the gas dynamics in
the disc, as well as a planet module that computes the orbital
trajectory of the planet that migrates due to interaction with
 the disc. In our
model, the planet and disc only interact gravitationally.  The
disc is assumed to have negligible self-gravity.  We investigate
this problem in global 3D cylindrical $\mathbf{r} = (r, \phi, z)$
coordinates, implemented in a Godunov-type code that was developed
by our group \citep{KoldobaEtAl2016}.

\paragraph*{The equations of hydrodynamics.}
We model the evolution of the accretion disc using 3D equations of
hydrodynamics, which can be compactly written as
\begin{equation}
\parti{\vec{U}}{t} + \nabla \cdot \vec{F}(\vec{U}) = \vec{Q} ~,
\label{eq:hydro}
\end{equation}
where $\vec{U}$ is the vector of conserved variables and
$\vec{F}(\vec{U})$ is the vector of fluxes:
\begin{equation}
\vec{U} = \left[\rho, \rho S, \rho \vec{v} \right]^T, \quad
\vec{F}(\vec{U}) = \left[\rho \vec{v}, \rho \vec{v} S, \tensor{M}
\right]^T, \label{eq:hydro_state}
\end{equation}
and $\vec{Q} = [0, 0, -\rho \nabla \Phi]$ is the vector of source
terms. In the above equations, $\rho$ is the fluid density, $S
\equiv p/\rho^\gamma$ is the function of the specific entropy
(entropy per unit mass, which we will simply call entropy in this
paper)
 with $\gamma = 5/3$, $\vec{v}$ is the
velocity vector, $\Phi$ is the gravitational potential of the
star-planet system (given later on in this section) and
$\tensor{M}$ is the momentum tensor, with components $M_{ij} =
\rho v_i v_j + \delta_{ij} p ~, $ where $\delta_{ij}$ is the
Kronecker delta function and $p$ is the fluid pressure. In our
code, we use the entropy balance equation instead of the full
energy equation, because in problems which we solve, the shock
waves are not expected, at least shock waves of large intensity
where we cannot neglect the energy dissipation. The advantage of
this approach is that in the entropy balance equation (compared
with the full energy equation) there are no terms which
significantly differ in their values \footnote{In the energy
equation, the main terms are gravitational energy and the kinetic
energy of the azimuthal motion, which can be much larger than the
internal energy and the energy of the poloidal motion in the
vicinity of the gravitational center.}.

We include a viscosity term, with the  viscosity coefficient in
the form of $\alpha-$viscosity, $\nu_{\rm vis}=\alpha c_s H$.   In
most of our simulations we take $\alpha=0$, so that the disc does
not change its shape during the simulation runs \footnote{We test
the migration of planets in discs with different values of
$\alpha$
in Appendix \ref{sec:appen-viscosity}.}.

\paragraph*{Migration of a planet.}
To calculate the migration of a planet,  we first calculate the
gravitational potential of the star-planet system, which is given
by
\begin{equation}
\Phi = -\frac{G M_*}{|\mathbf{x}|} - \frac{G
M_p}{(|\mathbf{x}-\mathbf{x}_p|^2 + \epsilon^2)^{1/2}} + \frac{G
M_p}{|\mathbf{x}_p|^3} \mathbf{x} \cdot \mathbf{x}_p ,
\label{eq:grav_potl}
\end{equation}
where $\epsilon$ is a smoothing radius that is added to prevent
divergence at the planet's location (e.g.
\citealt{NelsonEtAl2000}). In 3D simulations there are no specific
physical restrictions to this value (compared with the 2D
simulations, e.g. \citealt{MuellerEtAl2012}), so that any small
value can be taken at which the numerical solution near the planet
is stable. However, the smallest possible value is desirable,
because a large smoothing radius may decrease the value of the
corotation torques acting on the planet (e.g.
\citealt{Masset2002}). In our models, we took a value equal to two
grid cells at $r=1.5$: $\epsilon = 2 \Delta r \approx 0.025$. Test
simulations at smaller and larger values of $\epsilon$ have shown
that the solution is stable at this value of the smoothing
parameter. The total mass of the disc is small compared to the
mass of the star,
and hence we neglect the contribution to the potential from the
disc matter.

Next, we calculate the force per unit mass acting on the disc,
$\mathbf{f} = -\nabla\Phi$, or:
\begin{equation}
\mathbf{f} = -\frac{GM_*}{|\mathbf{x}|^3}\mathbf{x} -
\frac{GM_p}{(|\mathbf{x}-\mathbf{x_p}|^2+\epsilon^2)^{3/2}}
\mathbf{(x-x_p)} - \frac{GM_p}{|\mathbf{x}_p|^3} \mathbf{x_p}  ~.
\label{eq:grav_accel}
\end{equation}
The first and second terms represent the gravitational forces from
the star and the planet, respectively. The final term accounts for
the fact that the coordinate system is centered on the star, which
is a non-inertial frame due to the presence of the companion
planet.


\paragraph*{Numerical method.}
To solve the hydrodynamic equations, we use the Godunov-type code
developed in our group
for solving different hydro- and magnetohydrodynamic problems in
astrophysics and for modelling migration of planets in accretion
discs \citep{KoldobaEtAl2016}. In this paper, we use the
hydrodynamic, three-dimensional version of the code.

Our code is similar in many respects to other codes which use
Godunov-type method, such as \textit{PLUTO}
\citep{MignoneEtAl2007}, \textit{FLASH} \citep{FryxellEtAl2000}
and \textit{ATHENA} \citep{StoneEtAl2008}. We solve the entropy
conservation equations in place of the full energy equation.

To calculate the migration of a planet, we use earlier developed
approaches (e.g. \citealt{Kley1998,Masset2000}). Namely, we
calculate the potential and forces acting from the planet onto the
disc and calculate the position of a planet using the value of
this force with an opposite sign (see also
\citealt{KleyNelson2012}).

We rotate our numerical grid with a planet and the position of the
planet (relative to the grid) is fixed. Such approach has been
suggested by \citet{Kley1998} and has been used in a number of
planetary codes. In this approach, matter flow near the planet can
be calculated with a higher accuracy compared to cases of
non-rotating grids. In many recent codes, an orbital advection
scheme (the \textit{FARGO} algorithm), is used, where the layers
of the numerical grid have differential rotation (see
\citealt{Masset2000} and \citealt{Bentez-LlambayMasset2016} for 2D
and 3D versions of the \textit{FARGO} code) \footnote{This
algorithm helps to significantly speed up simulations compared
with cases of non-rotating and solidly-rotating grids (see also
\citealt{MignoneEtAl2012,McNallyEtAl2017}).}.

The equations of hydrodynamics are integrated numerically using an
explicit conservative Godunov-type numerical scheme
\citep{KoldobaEtAl2016}. In our numerical code the dynamical
variables are determined in the ``centers" of cells. For
calculation of fluxes between the cells we use the HLLD Riemann's
solver developed by \citet{MiyoshiKusano2005} and modified for
using the equation for the entropy balance (see also
\citealt{MiyoshiEtAl2010}) and grid rotation.
For better spatial resolution, we performed reconstruction of the
primitive variables to the boundaries between calculated cells
with the help of second-order limiter. Integration of the
equations with time is performed with a two-step Runge-Kutta
method. The code is parallelized using MPI.

\paragraph*{Boundary conditions.}
We use ``free" boundary conditions $\partial A/\partial r =0$ and
$\partial A/\partial z =0$ for all variables $A$. We forbid the
inward flow of matter.  We also use the procedure of damping waves
at the inner and outer boundaries, following
\citet{FromangEtAl2005}. Namely, we set the buffer zone for
damping at the inner part of the disc:
 $r_{\rm in} < r < 1.375 r_{\rm in}$
and at the outer part of the disc: $0.8 r_{\rm out}< r < r_{\rm
out}$.

\paragraph*{The grid.}
We use a 3D grid in cylindrical coordinates $(r, \phi, z)$. The
grid is centered on the star. In the radial direction, the inner
and outer boundaries are located at $\tr_{\rm in}=0.4$ and
$\tr_{\rm out}=5.2$, respectively. The grid resolution is set to
be higher in the inner region of the disc, where the grid is
compressed and evenly spaced in the interval of $0.4<r<2.6$, and
the number of grid cells is $N_r = 176$. Outside of this region,
the grid is spaced geometrically, so that the total number of grid
cells in the radial direction is  $N_r = 232$. The grid is evenly
spaced in the azimuthal and vertical directions, where the number
of grid cells are $N_\phi=480$ and $N_z=80$, respectively. In
experimental models with thicker discs, $h=0.05$ and $h=0.1$, the
number of grid cells in the vertical direction is $N_z=160$ and
$N_z=320$, respectively (see Appendix \ref{sec:appen-aspect
ratio}).

\begin{figure}
\centering
\includegraphics[width=0.5\textwidth]{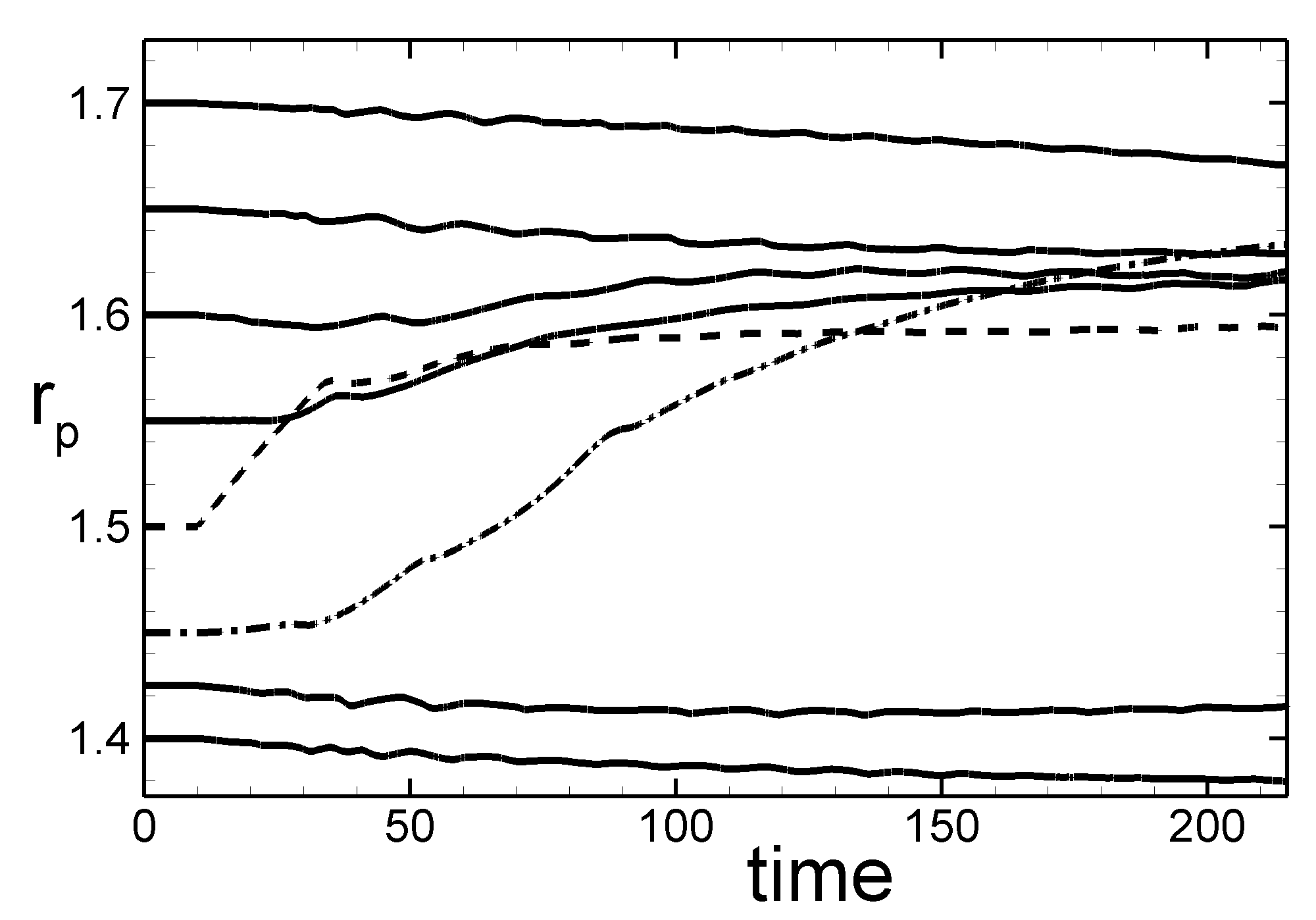}
\caption{Migration tracks for planets of mass $M_p=5 M_\oplus$ and
    initial orbital radii in the range of $r_p$ = 1.4--1.7. Tracks that start at
    $r_p=1.45$ and $r_p=1.5$ are shown as
    dashed and dash-dotted lines.} \label{fig:rp-5mp}
\end{figure}

\section{Results} \label{sec:migration-boundary}

Initially,  the transition between the disc and cavity is very
sharp (see top panel of Fig. \ref{fig:xz}). However, our
equilibrium is not precise, and the density in the disc has been
slightly redistributed during the first 10-20 rotations.
The density redistribution led to a smoother distribution at the
disc-cavity boundary (see middle panel of the same figure). Later
on, the density distribution in the disc varied only slightly over
time (see the middle and bottom panels corresponding to $t=20$ and
$t=100$, respectively). The right panel of Fig. \ref{fig:xz} shows
the linear distribution of the $\phi-$averaged surface density at
$t=0, 20, 50$ and $100$.

The planet is initialized on a fixed circular orbit; the disc is
allowed to relax from \ttime\ = 0 to $t=10$ before the planet is
released and permitted to migrate. We place a planet in different
parts of the disc, in the cavity and in the disc-cavity boundary,
and investigate its migration.

\subsection{Migration in the inner disc}
\label{subsec:migration-disc}

First, as a test case, we model the migration of planets in the
inner disc, where the surface density is approximately constant.
We placed a planet at an initial radius of $r_p=2$, away from the
disc-cavity boundary, and observed a slow inward migration towards
the star.

The migration path of a planet is mainly determined by the
Lindblad and corotation torques
\citep{GoldreichTremaine1979,GoldreichTremaine1980,Ward1986,Artymowicz1993a,Ward1997}.
The Lindblad torque is generated when a planet excites $m$-armed
waves in the disc with ``orbital'' frequencies $\omega =
m\Omega_{\rm p}$. The dispersion relation for these waves in a
low-temperature, low-mass disc is $\left(\omega -
m\Omega\right)^{2} = \kappa^{2},$ where $\kappa = \Omega_{\rm
Kep}$ is the epicyclic frequency in a Keplerian disc. Substituting
$\Omega(r)~=~\sqrt{{GM_{\star}}/{{r^{3}}}}$, the locations of the
Lindblad resonances are
 \footnote{Note that
Eq. \ref{eq:Lindblad} accurately describes the position of the
Lindblad resonances if the value of $\xi=m(H/r)<<1$
 \citep{GoldreichTremaine1980,Artymowicz1993a}. In our main model, where $H/r=0.03$,
we obtain this condition in the form: $m<<33$.}:
\begin{equation}
    r_{\rm LR} = r_{\rm p} \left(\frac{m \pm 1}{m} \right)^{2/3}.
    \label{eq:Lindblad}
\end{equation}

Fig. \ref{fig:schematic} shows that a planet generates two spiral
density waves in the disc. It also shows  the position of the
low-order Lindblad resonances. The lowest order outer Lindblad
resonances (OLR) are located at $r_{\rm OLR,m=1}\approx 1.59
r_p\approx 3.18$, $r_{\rm OLR,m=2}\approx 1.31 r_p\approx 2.62$,
$r_{\rm OLR,m=3}\approx 1.21 r_p\approx 2.42$, etc. All of them
are located in the disc and exert a negative torque on the planet.
 The lowest-order inner
Lindblad resonance (ILR),  $r_{\rm ILR,m=2}\approx 0.63 r_p\approx
1.26$, is located inside the low-density cavity, while the higher
order resonances, $r_{\rm ILR,m=3}\approx 0.76 r_p\approx 1.52$,
$r_{\rm ILR,m=4}\approx 0.82 r_p\approx 1.65$, etc., are located
at the disc-cavity boundary and in the disc. They exert a positive
torque on the planet. The torque from the OLR is larger in
magnitude, causing the overall cumulative torque to be negative.
 This drives the inward migration of the planet. Note
 that the torque per unit radius increases with $m$, and the
 cumulative torque is determined by the Lindblad resonances with
 high $m-$numbers
(e.g. \citealt{GoldreichTremaine1980}).

\begin{figure*}
\centering
\includegraphics[width=0.7\textwidth]{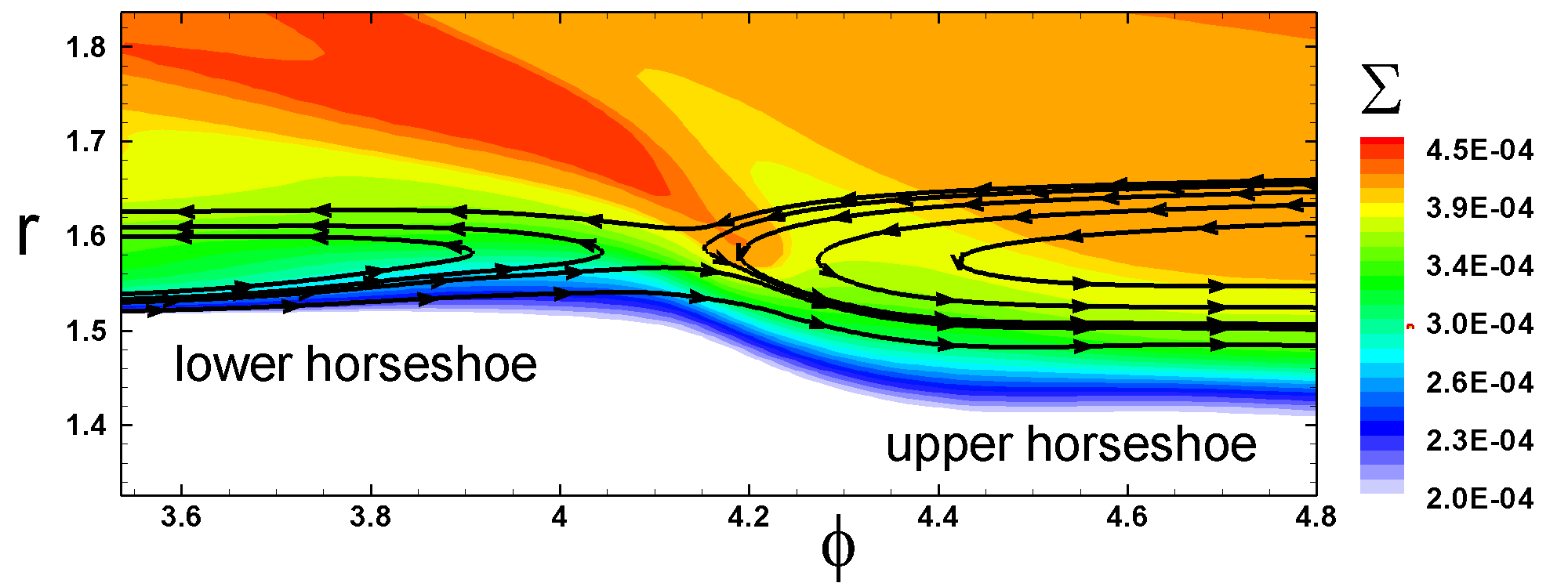}
\caption{The horseshoe orbits of gas around a planet of mass $15
M_\oplus$, located at the disc-cavity boundary. The colour
background shows surface density.} \label{fig:stream-1}
\end{figure*}

The corotation torque arises from the material in the planet's
co-orbital region where $\Omega \approx \Omega_p$.  The physics of
the corotation torque and its effect on planet migration have been
studied by a number of authors both theoretically and numerically
(e.g.
\citealt{GoldreichTremaine1979,Ward1991,OgilvieLubow2003,OgilvieLubow2006,MassetOgilvie2004,PaardekooperMellema2006,BaruteauMasset2008,PaardekooperPapaloizou2009a,
KleyEtAl2009,MassetCasoli2009,MassetCasoli2010,DAngeloLubow2010,PaardekooperEtAl2010,PaardekooperEtAl2011,Paardekooper2014}).

The co-orbiting material undergoes a horseshoe-shaped orbits in
the vicinity of the planet and asymmetries in the corotation
region lead to exchange of angular momentum between the planet and
the disc matter (e.g. \citealt{Ward1991}).
\citet{GoldreichTremaine1979} used the linear approximation and
calculated the corotation torque as a superposition of torques
from waves excited in the corotation region. They have shown that
the value and the sign of the corotation torque strongly depend on
the sign of the local density gradient.

\citet{TanakaEtAl2002} used the linear approximation to calculate
both the Lindblad and the corotation torques  in 2D and 3D
isothermal discs, and obtained the total torques as follows:
\begin{eqnarray}
\noindent\Gamma_{\rm (L+C),3D} = -(1.364+0.541s) \Gamma_0~, \\
\label{eqn:gamma-Tanaka-3D} \Gamma_{\rm (L+C),2D} =
-(1.160+2.828s) \Gamma_0~, \label{eqn:gamma-Tanaka-2D}
\end{eqnarray}
where $\Gamma_0=({q_p}/{h})^2 \Sigma_p r_p^4\Omega_p^2$.
Here,  $\Sigma_p\sim r^{-s}$ is the surface density in the disc at
the orbital radius of the planet, $r=r_p$. In 3D isothermal discs,
the total torque becomes positive (and the planet migrates
outward) if $(1.364+0.541s) < 0$, that is, at condition $s <
-2.52$, which corresponds to a high positive density gradient. In
2D discs, the condition is $s<-0.41$. In our disc, the surface
density distribution is flat, $s=0$, the total torque is negative,
and the planet migrates inward. Overall, we observed typical type
I migration similar to that observed by other authors who
performed 2D or 3D simulations (see, e.g., review paper by
\citealt{KleyNelson2012} and references therein).

\subsection{Migration at the disc-cavity boundary}

Next, we study the migration of planets at the disc-cavity
boundary. For that, we place a planet
in different parts of the disc-cavity boundary and its vicinity
and study its migration.

\fig{fig:rp-5mp} shows the
 migration tracks of planets with mass $M_p=5 M_\oplus$, which start at
different radii $r_p$ from the star, ranging
 from $r_p=1.4$, which corresponds to the low-density cavity, to
 $r_p=1.7$, which corresponds to the inner part of the disc. This figure shows that the
planets starting just outside the disc-cavity boundary (at $r_p$ =
1.65 and  $r_p$ = 1.7) migrate inward due to differential Lindblad
torque, which is analogous to the migration in the disc discussed
in Sec. \ref{subsec:migration-disc}. The torque is mainly
determined by the OLR, because many of the ILR are located in the
low-density cavity. The torque is negative and the planets migrate
inward. Later on, they stop migrating because they reach the
regions of large positive density gradient, where the corotation
torque is high, and their migration stops due to the balance of
the torques.

Planets that start migrating from a region of high density
gradient ($r_p=1.45, 1.5, 1.55$) migrate rapidly outward. Later
on, when they reach the parts of the disc with a lower density
gradient, their migration slows down and stalls at the radius
where the positive corotation torque is balanced by the negative
Lindblad torque.

\begin{figure*}%
\centering
\includegraphics[width=0.9\textwidth]{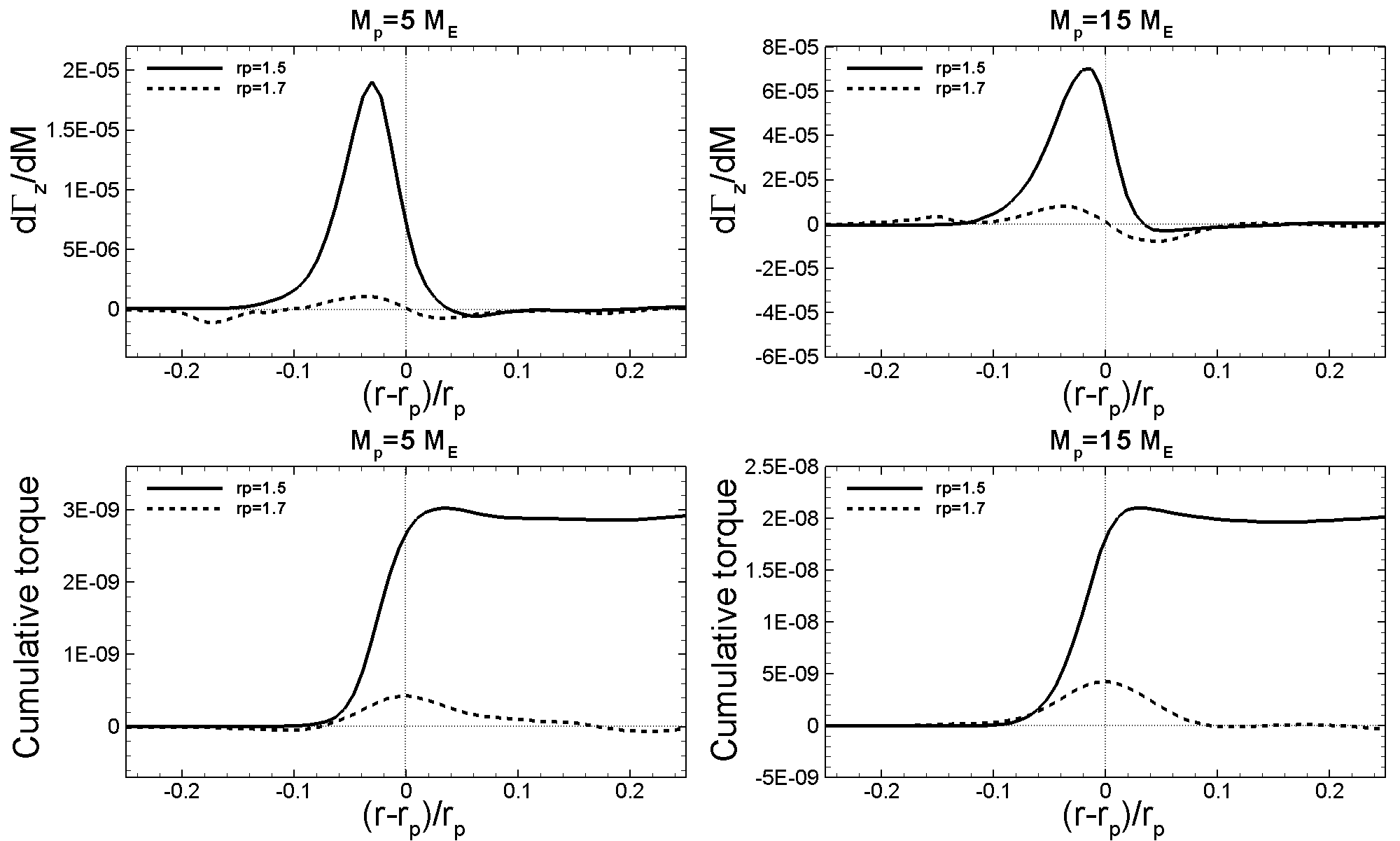}
\caption{Torque profiles. {\it Top panels:} torque per unit disc
mass for
planets with masses  $5 M_\oplus$ and $15 M_\oplus$, located at
distances  $r_p=1.5$ and $1.7$ from the star, at $t = 20$. {\it
Bottom panels:} cumulative torque as a function of normalized
radius.} \label{fig:dTdM-m5-15}
\end{figure*}

Fig. \ref{fig:rp-5mp} shows that the orbits of the planets
migrating from smaller and larger radii converge at the radius of
$r_{\rm trap}\approx 1.6$, which corresponds to the planet trap
radius. This analysis shows that there is a stable halting region
where the positive corotation torque  balances the negative
differential Lindblad torque, and therefore the planet is trapped.

In another set of numerical experiments, we studied the migration
of planets with mass $M_p=15M_\oplus$ and observed a similar
convergence of migration paths at the radius of $r_{\rm
trap}\approx 1.65$. However, the planets migrated approximately
three times faster, which is in agreement with the dependence for
the migration time scale, $\tau\sim M_p^{-1}$, derived from
theoretical studies (see, e.g. Eq. 70 of
\citealt{TanakaEtAl2002}).

The corotation torque is associated with the asymmetry of the
horseshoe shaped orbits of matter near a planet \citep{Ward1991}.
Namely, in the coordinate system of the planet, the co-orbital
matter performs U-turns near the planet, such that the matter in
front of the planet  transfers its angular momentum to the planet,
while the matter behind the planet takes angular momentum away
from the planet.

Fig. \ref{fig:stream-1} shows an example of the strong asymmetry
of the horseshoe orbits around a planet with mass $M_p=15
M_\oplus$, located in the region of the high density gradient
($r_p\approx 1.5$). One can see that matter in the upper horseshoe
(located at the right side of the plot) carries more mass (and
angular momentum) than the lower horseshoe (left side of the
plot). Matter comes to the upper horseshoe orbits from the
high-density disc (located above the planet in the plot), while
matter to the lower horseshoe comes from the low-density cavity
(located below the planet in the plot). The positive torque
associated with the upper horseshoe is much larger than the
negative torque associated with the lower horseshoe, and the
planet migrates outward.

To investigate the asymmetry of the torques acting at the
disc-cavity boundary, we calculated the torque per unit disc mass
(e.g. \citealt{MassetEtAl2006a,DAngeloLubow2008})\footnote{Since
the aim of this work is to study migration, we consider only the
$z$ component of the torque, which tracks the forces in the disc
plane.}:
\begin{equation}
\frac{d\Gamma_z}{dM} = \frac{1}{2 \pi \Sigma_{\rm avg}}
\int_{-\infty}^{\infty} \int_0^{2 \pi} \gamma_z d\phi dz ~
\label{eq:torque_per_mass}
\end{equation}
as a function of normalized radius $(r-r_p)/r_p$, where
\begin{equation}\gamma_z = \rho(\mathbf{x}) \frac{\partial \Phi_p}{\partial
\phi} = \frac{GM_p\rho(\mathbf{x})}{(|\mathbf{x}-\mathbf{x_p}|^2 +
\epsilon^2)^{3/2}} r r_p \sin(\phi-\phi_p) ~
\end{equation}
is the torque per unit volume (torque density) acting on the
planet from the disc, and  $\Phi_p$ is the gravitational potential
of the planet (second term in \eqn{eq:grav_potl}). The top panels
of \fig{fig:dTdM-m5-15} show torque per unit mass for planets with
masses $M_p=5 M_\oplus$ and $M_p=15 M_\oplus$. The bottom panels
of the same figure show the cumulative torque, i.e., the torque
density integrated over a given radius.

Fig. \ref{fig:dTdM-m5-15} shows that, for a planet that migrates
from radius $r_p = 1.7$ (dashed line),
 both the inner and outer Lindblad resonances contribute to
the torque on the planet. The total torque from the OLR is larger
in magnitude, causing the cumulative torque to be negative. This
drives inward migration of the planet, as shown in
\fig{fig:rp-5mp}. For a planet at $r_p = 1.5$, there is a strong
positive torque from the region interior to the planet, resulting
in a large positive cumulative torque (bottom panels). This large,
positive torque causes the planet to migrate rapidly outward.

\begin{figure*}
    \centering
   \includegraphics[width=0.9\textwidth]{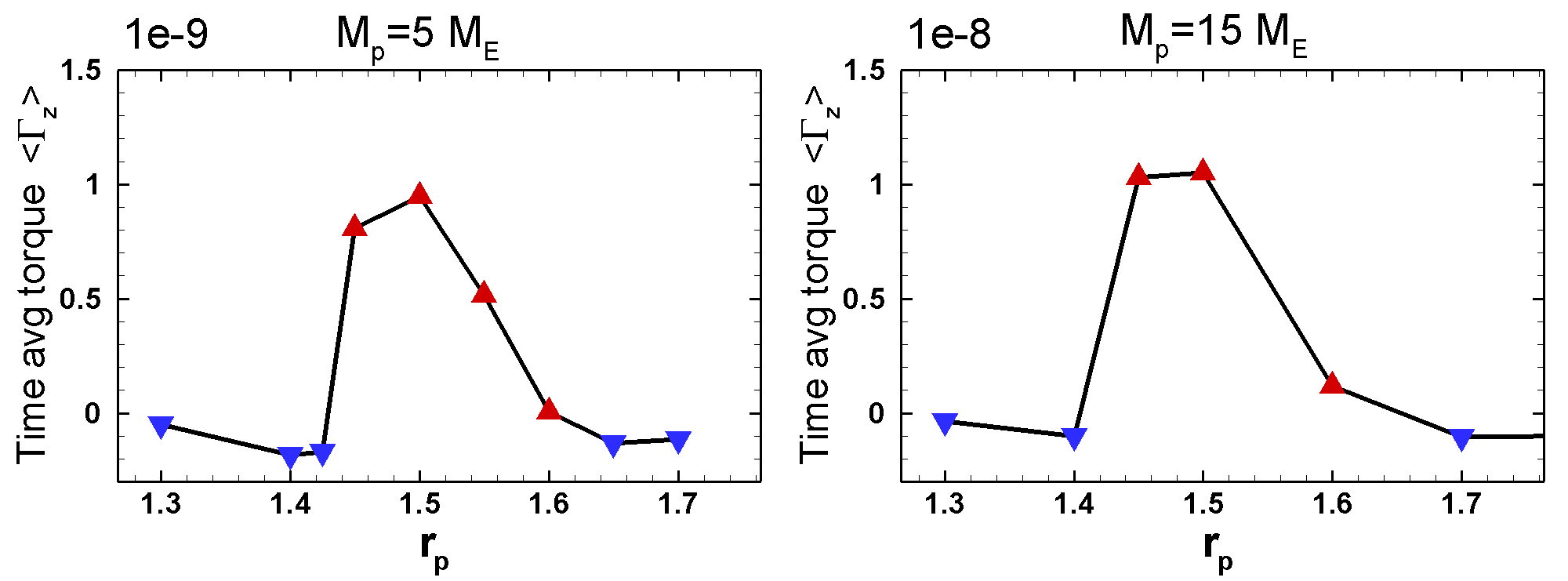}
    \caption{The time-averaged total torque. The plot shows $\left<\Gamma^{\rm tot}_z\right>$ versus planet radius $r_p$
    for the planets with initial orbits near the disc-cavity boundary.
    The torques are computed by time averaging between $t$ = 10 (when the planet is first allowed to migrate) and $t$ = 60.
     Left and right panels show the torques for planets with masses
    $M_p=5 M_\oplus$ and $M_p=15 M_\oplus$, respectively.} 
    \label{fig:torque-av-5-15}
\end{figure*}

The degree of asymmetry between the torques sets the direction of
the planet's overall migration. This net torque can be measured by
integrating the torque density over the simulation region:
\begin{equation}
\Gamma_z = \int_\mathcal{V}{\gamma_z} dV = GM_p
\int_\mathcal{V}{\rho(\mathbf{x}) \frac{r r_p
\sin(\phi-\phi_p)}{(|\mathbf{x}-\mathbf{x_p}|^2 +
\epsilon^2)^{3/2}} dV} ~. \label{eq:total torque}
\end{equation}
We calculated the time-averaged torque (during the time interval
of $t=10-60$) acting on the planets located at different distances
$r_p$ from the star. \fig{fig:torque-av-5-15} shows that the
positive torque on the planets at the disc-cavity boundary  (red
triangles) is much larger than the negative torque acting in the
disc or cavity. This contrast shows how powerful the trapping
mechanism is: the planets that try to migrate into the cavity
enter the region of a high positive density gradient and
experience strong positive torque, which prevents them from
migrating. One can see that the torque is zero at $r_{\rm trap}
\approx 1.6$ and $r_{\rm trap} \approx 1.65$ for planets with
masses $M_p=5 M_\oplus$ and $M_p=15 M_\oplus$, respectively.
Planets in these regions are stably trapped, because the torque
interior to $r_{\rm trap}$ is positive and drives outward
migration towards the trap; similarly, the torque exterior to
$r_{\rm trap}$ is negative and also drives inward migration
towards the trap. Overall, these 3D simulations confirm the
results of the earlier studies  by  \citet{MassetEtAl2006a} and
\citet{MorbidelliEtAl2008}, who predicted the existence of stable
planet traps in their 2D simulations. Below, we compare results of
our 3D simulations with results obtained in theoretical studies
and 2D simulations.

\subsection{Non-linear horseshoe drag and the width of the
horseshoe region}

 The corotation torque associated
with horseshoe orbits has been calculated in general, non-linear
approach by \citet{Ward1991} in 2D isothermal discs as:
\begin{equation}
\Gamma_{\rm HS}=\frac{3}{4}\Sigma_p \Omega_p^2 x_s^4 \bigg[
\frac{d{\rm log}(\Sigma/B)}{d {\rm log}r}\bigg]~,
\label{eq:Gamma-HS}
\end{equation}
where $B=(1/2r)d(r^2\Omega)/dr$ is vorticity, and $B/\Sigma$ is
specific vorticity (or vortencity). The torque depends strongly on
the half-width of the horseshoe region, $x_s$. For low-mass
planets (where $q_p<<h^3$) located in isothermal discs, $x_s$ has
been estimated as:
\begin{equation}
x_s\approx k_s r_p\sqrt{q_p/h}~, \label{eqn:x_s}
\end{equation}
where slightly different values of $k_s$ were found in different
studies: $k_s=0.96$ \citep{Masset2001} and $k_s=1.16$
\citep{MassetEtAl2006b,PaardekooperPapaloizou2009b}.

\begin{figure*}
\centering
\includegraphics[width=1.0\textwidth]{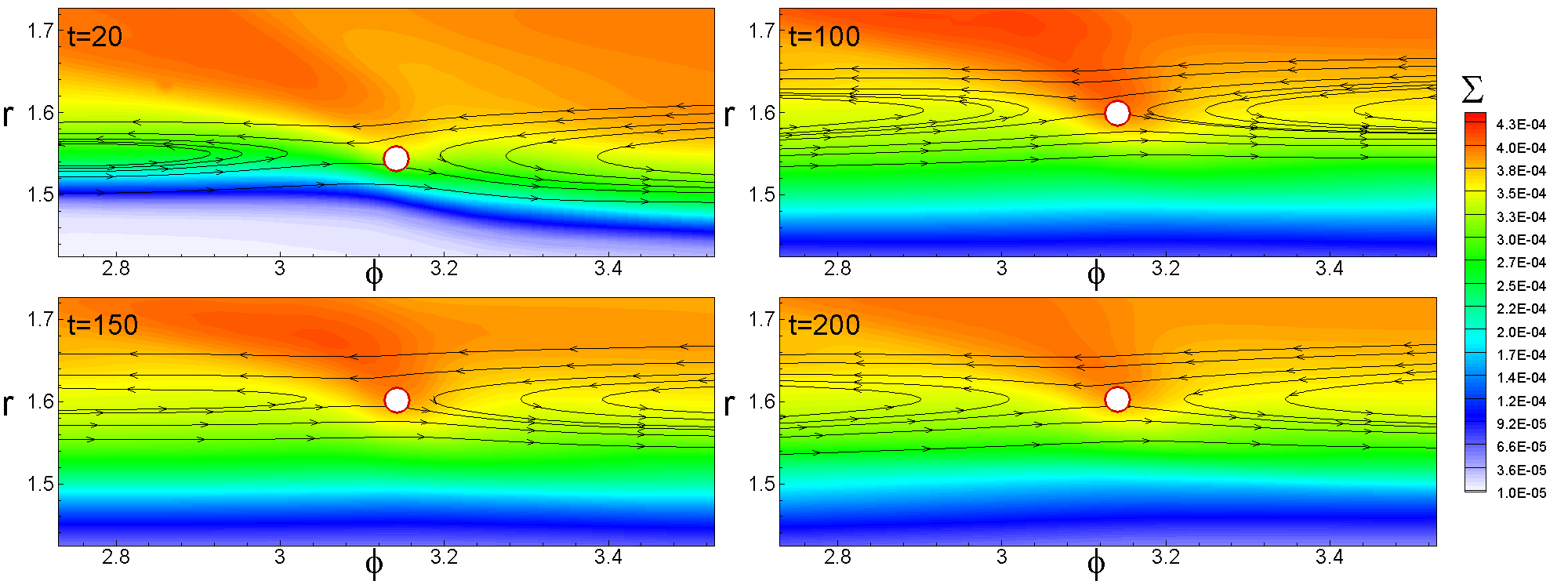}
\caption{Streamlines of matter flow around a planet of mass $5
M_\oplus$ with initial location at $r_p=1.5$ at $t=20, 100, 150,
200$. The colour background shows the surface density, $\Sigma$.
The circle shows the position of the planet (not to scale).}
\label{fig:stream-mp5-4}
\end{figure*}

We can find the values of $x_s$ and $k_s$ from our simulations.
Fig. \ref{fig:stream-mp5-4} shows streamlines of matter flow
around a planet of mass $5 M_\oplus$, with its initial location at
$r_p=1.5$, at different moments in time: $t=20, 100, 150, 200$.
Initially, at $t=20$, the planet is located at the steep part of
the density slope, and the horseshoe orbits are strongly
asymmetric. Later on, a planet moved to larger radius, $r_p\approx
1.6$, where the density gradient is not as high, and orbits become
somewhat more symmetric. However, in all cases the upper and lower
horseshoe orbits originate in the regions of high and low density,
respectively, which provides asymmetry in the corotation torque.
From Fig. \ref{fig:stream-mp5-4}, we find the approximate value of
the observed width of the horseshoe region, $\Delta r\approx 0.11$
and obtain $x_s=0.5\Delta r/r_p\approx 0.034$. Using this value,
and taking $r_p=1.6$, $q_p=1.5\times10^{-5}$ and $h=0.03$, we
obtain $k_s\approx 0.95$. This value is in agreement with earlier
findings
\footnote{Note that our value of $x_s$ is somewhat smaller than
the value $k_s=1.16$ obtained from most recent 2D simulations
(e.g. \citealt{MassetEtAl2006b,PaardekooperPapaloizou2009b}). The
half-width of the horseshoe orbits found by
\citet{MassetBenitez2016} and \citet{LegaEtAl2015} in their 3D
simulations is about 10\% smaller than that observed in 2D
simulations. }.

\subsection{Saturation of the corotation torque}

\label{subsec:saturation}

The horseshoe corotation torque is prone to saturation, because a
planet exchanges angular momentum with the co-orbital matter,
which cannot transfer the angular momentum to other parts of the
disc unless new matter enters the corotation region due to
viscosity or some other mechanism (e.g.
\citealt{Ward1991,Masset2001,MassetOgilvie2004,OgilvieLubow2006}).
Saturation will not occur if new matter enters the horseshoe
region with the time scale smaller than the libration time scale
\citep{Masset2001,Masset2002}:
\begin{equation}
\tau_{\rm lib}=\frac{8\pi}{3{\bar x}_s} \Omega_p^{-1} ~,
\label{eqn:libration time scale}
\end{equation}
during which a fluid element at the orbital radius $r_p(1+{\bar
x}_s)$ completes two orbits in the frame corotating with the
planet. The disc turbulence, which provides an effective
viscosity, can help to prevent the corotation torque from
saturation. In our model we switched the module of viscosity off,
which helped us to support the same density distribution at the
disc-magnetosphere boundary \footnote{We analyze the role of
viscosity in Appendix \ref{sec:appen-viscosity}.}.

\begin{figure} 
\centering
    \includegraphics[width=0.5\textwidth]{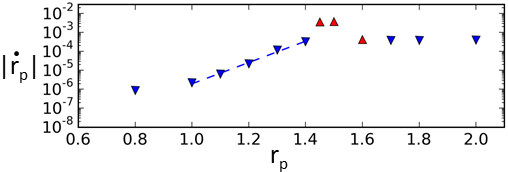}
    \caption{Absolute value of the migration rates of a planet of mass
$M_p=15M_\oplus$, with initial orbital radii ranging from $r_p$ =
0.8--2.0. The downward pointing (in blue) triangles indicate
inward
migration, while the upward pointing triangles (in red) indicate outward migration.} 
    \label{fig:migrate-3d}
\end{figure}

However, in our model, the saturation does not occur, because
planets migrate rapidly, so that new matter comes into the
horseshoe region due to the planet's migration (e.g.
\citealt{PaardekooperEtAl2010,Paardekooper2014}). The torque will
stay unsaturated, if the time scale of migration through the
region $2x_s$ is smaller than the libration time scale:
$2x_s/{\dot r}_p < \tau_{\rm lib}$. The corresponding migration
rate is:
\begin{equation}
\dot{r}_p \gtrsim 1.33\times
10^{-4}\bigg(\frac{r_p}{1.5}\bigg)^{1/2}\bigg(\frac{M_p}{5
M_\oplus}\bigg) \bigg( \frac{h}{0.03}\bigg)^{-1} ~.
\label{eq:saturation migration rate}
\end{equation}
In application to our model, we obtain: ${\dot r}_p \gtrsim
1.33\times 10^{-4}$ and $\dot{r}_p \gtrsim 4.0\times 10^{-4}$ for
planets with masses $M_p=5M_\oplus$ and $M_p=15M_\oplus$,
respectively \footnote{To obtain the migration rate in dimensional
units, one should multiply $\dot{r}_p$ by $v_0$ from Tab.
\ref{tab:units}. For example, for cavity located at $1$ AU we
obtain: $\dot{r}_p\times v_0\approx 3.6\times 10^{-3}$ km
s$^{-1}$.}. We compare this rate with the migration rate observed
in simulations. Fig. \ref{fig:migrate-3d} shows the migration
rates obtained from simulations for planets with mass
$M_p=15M_\oplus$ \footnote{The migration rate has been estimated
by measuring the slope between the start and the midpoints of the
migration tracks.}. The measured migration rates at the
disc-cavity boundary, ranging as $\dot{r_p} = 3 \times 10^{-3} - 4
\times 10^{-4}$, are comparable or larger than the critical value
(obtained from Eq. \ref{eq:saturation migration rate}), and
therefore the corotation torque is not saturated.

When the planet reaches the position of zero torque ($r_p\approx
1.6-1.65$) the migration stalls, and the corotation torque can be
saturated. However, at the disc-cavity boundary, the conditions
for saturation may be different compared with other parts of the
disc. Namely, the processes which support the disc-cavity
boundary, will also support the asymmetry of the horseshoe orbits,
and would prevent the saturation of corotation torque.
 For example, in cases of the
disc-magnetosphere boundary,  the density in the disc is always
larger than the density in the cavity. The balance between the
matter pressure in the disc and  the magnetic pressure in the
cavity establishes rapidly, on a time scale comparable with the
Keplerian rotation at the boundary (e.g.
\citealt{RomanovaEtAl2003,BlinovaEtAl2016}), which is much shorter
than the libration time scale. Therefore, a sharp density gradient
and asymmetry of the horseshoe orbits would always be supported at
the boundary, and therefore high viscosity or rapid migration are
not required to prevent the corotation torque from saturation. In
cases of disc-cavity boundaries of different origins, the details
of saturation depend on the physics of these boundaries and should
be studied separately.

\begin{figure*}
\centering
\includegraphics[width=1.0\textwidth]{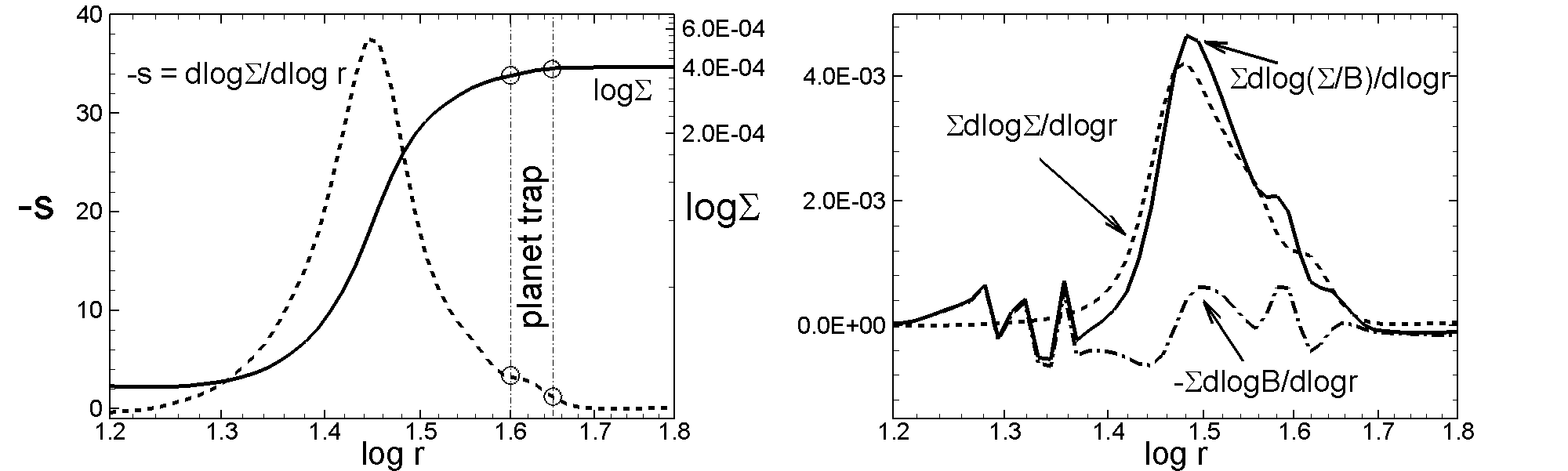}
\caption{\textit{Left panel:} The radial distribution of the
azimuthally-averaged surface density, $\Sigma$ (solid line), and
the power in the density distribution, $-s=d{\rm log}\Sigma/d{\rm
log} r$. Vertical lines and small circles show the range of
parameters at which the low-mass planets are trapped.
\textit{Right panel:} Components of the horseshoe torque
associated with the density gradient (dashed line), with the
gradient of vorticity, $B$ (dash-dot line), and the total torque
(solid line).} \label{fig:1d-HS-terms}
\end{figure*}

\subsection{Comparisons with 2D simulations of planet trapping}
\label{subsec:comparison-Masset}

\citet{MassetEtAl2006a} considered the migration of low-mass
planets of $1-15 M_\oplus$ at the disc-cavity boundary,
where the disc and the cavity have uniform
surface densities of $\Sigma_o$ and $\Sigma_i$, respectively. They
supported this configuration, placing different values of
$\alpha-$parameter of viscosity in the disc and in the cavity
($\alpha=10^{-4}$ and $\alpha=10^{-2}$, respectively), which
helped them to evacuate matter from the cavity in steady-state
fashion. Compared with their simulations, we consider inviscid
discs, where the quasi-steady disc-cavity configuration
is supported for a long time due to very low (numerical)
viscosity.

\citet{MassetEtAl2006a} considered the disc-cavity boundaries with
the width  of the transition region, $\lambda\approx 4H<<r$, and
density ratios between the disc and the cavity of
$\Sigma_d/\Sigma_{\rm cav}=7$ and $1.4$. In our simulations, we
have similar width of the transition region, $\lambda\approx
0.2\approx 4.4 H$, but larger density ratio between the disc and
the cavity, which is $\rho_d/\rho_{\rm cav}=100$ in the equatorial
plane, which corresponds to $\Sigma_d/\Sigma_{\rm cav}\approx 14$
for surface densities.

\citet{MassetEtAl2006a} noted that, in the transition region, the
term $d \rm{log}\Sigma/d \rm{log}r$ is a sharply peaked function
with a maximum that scales as $|{d \rm{log}\Sigma}/{d \rm{log}
r}|_{\rm max} \sim {r}/{\lambda}\rm{log}({\Sigma_o}/{\Sigma_i})~ .
$ The left panel of Fig. \ref{fig:1d-HS-terms} shows typical
surface density distribution in our model and the distribution of
the power
 $-s=d\rm{log}\Sigma/d \rm{log}r$ with radius. One can see
that $-s$ has small values at the edges of the disc-cavity
boundary, but is sharply peaked at $-s\approx 36$ in the middle of
the boundary, at $r\approx 1.45$.

 \citet{MassetEtAl2006a} analyzed the radial distribution of the
main terms in  the horseshoe corotation torque,
$\Gamma_C\sim\Sigma{d\rm{log}(\Sigma/B)}/{d{\rm log} r}~$:
\begin{equation}
\Sigma\frac{d\rm{log}(\Sigma/B)}{d{\rm log} r} =
\Sigma\frac{d\rm{log}\Sigma}{d{\rm log} r} - \Sigma\frac{d\rm{log}
B}{d{\rm log} r} ~, \label{eq:Gamma-HS-terms}
\end{equation}
where the left and right terms are associated with  the surface
density gradient and the vorticity gradient, respectively.
\citet{MassetEtAl2006a} compared these two terms and concluded
that they are comparable and both contribute to the corotation
torque (see Fig. 1 from \citealt{MassetEtAl2006a}).

We also calculated and compared these terms, taking one of our
models, where a planet of mass $5M_\oplus$ started its migration
at $r_p=1.55$ and migrated outward. The right panel of Fig.
\ref{fig:1d-HS-terms} shows these terms separately and their sum
at the moment of time corresponding to $t=100$. One can see that
the density gradient term is much larger than the gradient of
vorticity term. This can be explained by the fact that in our
model the density gradient at the disc-cavity boundary is larger
than that in the model of \citet{MassetEtAl2006a}.

\citet{MassetEtAl2006a} observed that planets migrating from the
disc  halt their migration at the edge of the disc, before
entering the low-density cavity.  In our simulations we also
observed that the migration halts at the disc edge, before
entering the sharp boundary, though we considered the migration
from either side, from the disc or from inside the boundary.

From the left panel of Fig. \ref{fig:1d-HS-terms}, we see that the
trapping region is located in parts of the disc with relatively
small positive density gradients. Left and right vertical lines
show the positions of the planet trap, $r_{\rm trap}\approx 1.6$
and $r_{\rm trap}\approx 1.65$, for planets with masses of
$5M_\oplus$ and $15M_\oplus$, respectively. The density gradients
are $-s=3.3$ and $-s=1.2$, respectively.
 These values approximately correspond to the value of $s\approx -2.52$
for zero torque in Eq. \ref{eqn:gamma-Tanaka-3D} from
\citet{TanakaEtAl2002} for 3D discs \footnote{Note that
\citet{TanakaEtAl2002} considered isothermal discs, while our
model is adiabatic. See Sec. \ref{subsec:comparisons-adiabatic}
for comparisons of our model with 2D adiabatic models.}.

Note that a large density drop is not necessary for planet
trapping:  simulations show that planets are trapped at the parts
of the disc with relatively small values of the positive density
gradient (see also Sec. \ref{subsec:minimum-density}).

Overall, our 3D simulations confirm the trapping mechanism
proposed and studied by \citet{MassetEtAl2006a} in 2D simulations.
Our Fig. \ref{fig:torque-av-5-15} and Fig. 1 from
\citet{MassetEtAl2006a} show that the positive corotation torque
strongly increases when a planet moves away from its trapping
location towards the disc-cavity boundary, and therefore the
boundary represents a stable trap for the low-mass planets.

\subsection{Comparisons with 2D models of adiabatic discs}
\label{subsec:comparisons-adiabatic}

In our simulations, we used the adiabatic equation of state
\footnote{Note that \citet{MassetEtAl2006a} used
locally-isothermal equation of state.}. Below, we compare results
of our 3D simulations with 2D simulations of
\citet{PaardekooperEtAl2010}, who studied torques acting on the
low-mass planets in  case of non-isothermal, adiabatic discs. In
their 2D model, no density jumps were considered, but instead the
surface density, temperature and entropy smoothly vary with radius
as
\begin{equation}
\Sigma\sim r^{-s}, ~T\sim r^{-\beta}, ~ S\sim r^{-\xi ~} .
\end{equation}
From their simulations, they derived analytical dependencies for
the linear Lindblad torque and non-linear (horseshoe) corotation
torque (see their equations 14 and 45) as
\begin{eqnarray}
\gamma \Gamma_L/\Gamma_0 = -(2.5+1.7\beta-0.1s){\bar b}^{0.71}~,
\end{eqnarray}
\begin{eqnarray}
\gamma \Gamma_C/\Gamma_0 =  1.1{\bar b}\bigg(
\frac{3}{2}-s\bigg)+\frac{\xi}{\gamma}{\bar b}\bigg(10.1\sqrt{\bar
b}-2.2 \bigg)~, \label{eq:corotation-paardekooper}
\end{eqnarray}
where ${\bar b}=0.4/(b/h)$, $b$ is the softening parameter in the
equation for the potential of the star-planet system.  The left
and right terms in the right-hand side of eq.
\ref{eq:corotation-paardekooper} for the corotation torque are
connected with the vortencity and entropy gradients, respectively.

For comparisons with our model, we take the value of softening
$\epsilon=0.025$, used in our model, and obtain
$b=\epsilon/r_p=0.025/r_p\approx 0.0167$ (at $r_p=1.5$) and $\bar
b=0.4/(b/h)\approx 0.72$. We also take into account that in
adiabatic discs $\beta=\xi+(\gamma-1)s ~$  and take the value
$\gamma=5/3$ used in our model, and obtain the total torque in the
form:
\begin{eqnarray}
\gamma \Gamma_{(L+C)}/\Gamma_0 = -0.79 - 1.61 s + 1.40 \xi~.
\label{eq:corotation-paard-total}
\end{eqnarray}
In cases of steady cold discs (which are in the force equilibrium)
the pressure gradient force is small and pressure varies only
slowly with radius. Then from the ideal gas equation, one obtains
$\beta\approx -s$, and $\xi=\beta-(\gamma-1)s \approx -\gamma s$.
Substituting these values to Eq. \ref{eq:corotation-paard-total},
one obtains that the torque is positive, if $s \lesssim -0.2$.

\begin{figure}
\centering
\includegraphics[width=0.5\textwidth]{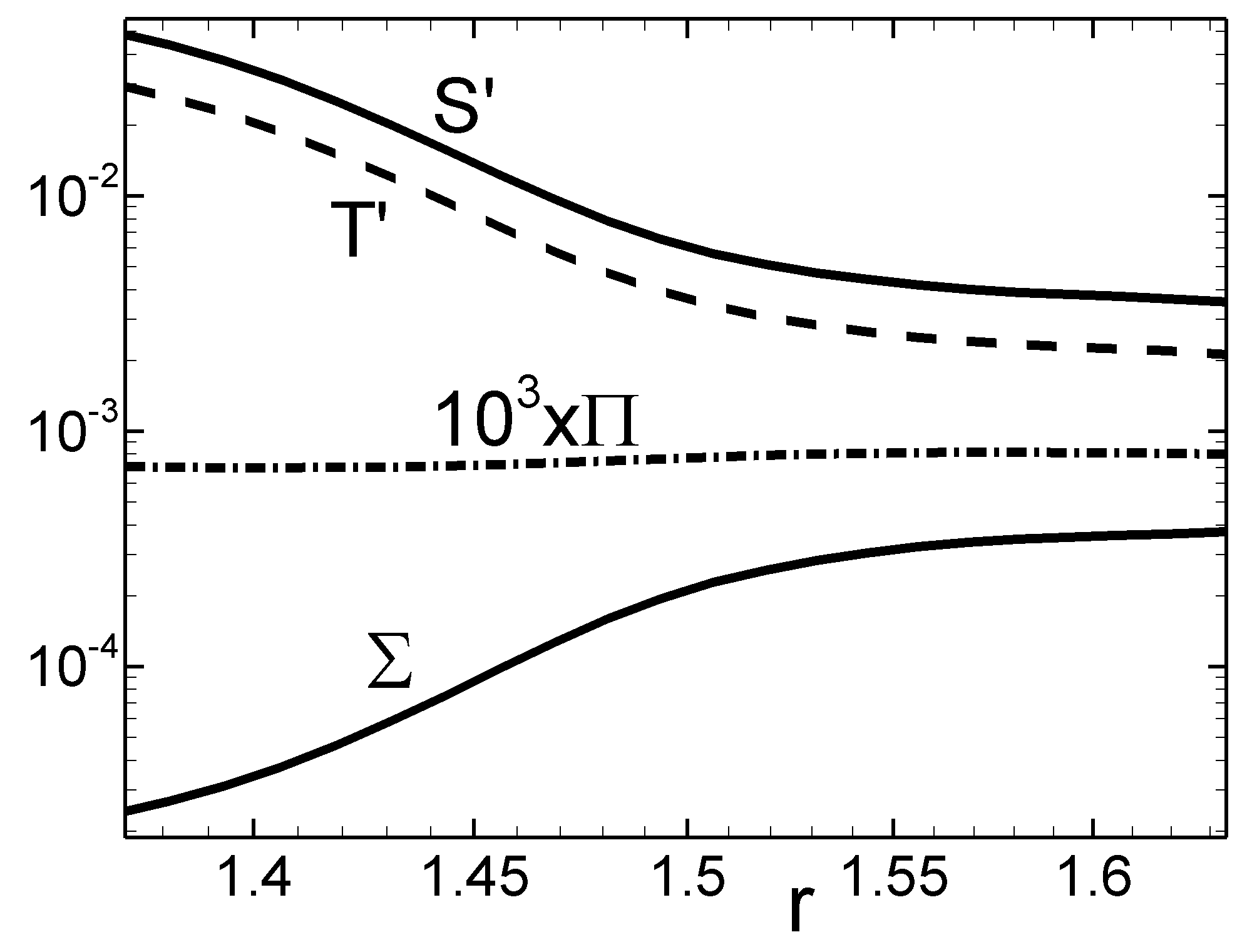}
\caption{The radial distribution of the azimuthally-averaged
values at the disc-cavity boundary at $t=100$: the surface
density, $\Sigma$ (solid line), surface pressure, $\Pi$ (dash-dot
line), surface entropy $S'$ (solid line), and surface temperature,
$T'$ (dashed line).} \label{fig:1d-sig-entr}
\end{figure}

To compare our 3D model  with the above 2D model, we integrate
different values obtained in our model in z-direction and obtain
surface density, $\Sigma$, surface pressure, $\Pi$, surface
temperature, $T'$ and surface entropy, $S'$. Fig.
\ref{fig:1d-sig-entr} shows typical radial distribution of these
values at  the disc-cavity boundary. One can see that the surface
pressure is almost constant, as expected, so that the quasi-steady
state is supported across the boundary. The surface entropy
$S'=\Pi/\Sigma^\gamma\approx \Sigma^{-\gamma}\sim r^{\gamma s}$
decreases with radius.  The surface temperature is connected with
pressure and density through an ideal gas law: $T'=\Pi/\Sigma$ (in
our dimensionless units), and is approximately inverse function of
the surface density, $T'\sim \Sigma^{-1}\sim r^s$.

Our simulations show that the migration of planets halts at radii
$r\approx 1.6-1.65$. From Fig. \ref{fig:1d-sig-entr} we derive the
slope in the surface density distribution in the interval of radii
of $1.6<r<1.65$ and obtain an approximate value of $s \approx
-(2.0-2.3)$. This value is larger than the value obtained from the
theoretical estimates, $s\approx - 0.2$. The difference may be due
to the fact that our simulations are three-dimensional, while the
comparisons were performed for two-dimensional discs.  For
example,  in cases of isothermal discs, \citet{TanakaEtAl2002}
obtained zero torque at $s\approx-0.41$ for 2D discs, and
$s\approx -2.52$ for 3D discs. We suggest that in cases of
adiabatic discs, the zero-torque value of $s$ can be also
different in 3D discs versus 2D discs.

\subsection{Minimum density drop for planet trapping}
\label{subsec:minimum-density}

In our model, the surface density ratio between the disc and
cavity is high, $\Sigma_d/\Sigma_{\rm cav}\approx 13.6$. Such a
high density ratio is not necessary for trapping. In this section
we estimate the minimum density ratio which is necessary for
trapping.

Our simulations show that planets are trapped at the edge of the
disc,  \textit{before} reaching the sharp density drop at the
disc-cavity boundary. They are trapped in parts of the disc where
the  positive density gradient $-s\approx 3.3$ for $M=5M_\oplus$
and $-s\approx 1.2$ for $M=15M_\oplus$. \citet{TanakaEtAl2002}
found from semi-analytical studies, that the positive corotation
torque balances the negative Lindblad torque (when a planet can be
trapped) at $-s=2.52$ (in the case of 3D discs).

To estimate the minimum density drop, which is needed for
trapping, we consider a cavity  with a linear distribution of the
surface density, which varies from $\Sigma_d$ in the disc at
$r=r_d$ to $\Sigma_{\rm cav}$ in the cavity at $r=r_{\rm cav}$.
Then we can write the power $s$ in the surface density slope as
\begin{equation}
 -s=\frac{d{\rm log}\Sigma}{d{\rm
log}r} = \frac{\Delta{\rm log}\Sigma}{\Delta{\rm log} r} =
\frac{{\rm log}(\Sigma_d/\Sigma_{\rm cav})}{{\rm log}(r_d/r_{\rm
cav})}~.
\end{equation}
One can see that the power $s$ depends on the density drop,
$\Sigma_d/\Sigma_{\rm cav}$, and also on the ratio $r_d/r_{\rm
cav}=1/(1-\Delta r/r_d)$, where $\Delta r=r_d-r_{\rm cav}$ is the
width of the transition region. Using these approximations, we can
estimate the density drop. For example, in our simulations
$r_d\approx 1.6$, $r_{\rm cav}\approx 1.4$,  $\Delta r \approx
0.2$, $\Delta r/r_d = 0.125$ and $r_d/r_{\rm cav}\approx 1.143$.
Then for values of $-s=1.2$ and $-s=3.3$ we obtain
$\Sigma_d/\Sigma_{\rm cav}\approx 1.2$ and $1.6$, respectively. If
we take $-s=2.52$ from \citet{TanakaEtAl2002} (see Eq.
\ref{eqn:gamma-Tanaka-3D}) then we obtain $\Sigma_d/\Sigma_{\rm
cav}\approx 1.4$. These estimates show that a small density drop
is sufficient for trapping low-mass planets. We note that
\citet{MassetEtAl2006a} tested the trapping region with the
density drop of $\Sigma_d/\Sigma_{\rm cav}\approx 1.4$ in their 2D
simulations and observed planet trapping.

\section{Applications} \label{sec:applic}

A planet can be trapped at the disc-cavity boundaries, which are
expected in several places of the protoplanetary discs.

\subsection{A planet trap at the disc-magnetosphere boundary}

Young low-mass stars at the centers of protoplanetary systems tend
to be highly magnetized due to the onset of convection in their
protostellar cores (e.g. \citealt{DonatiLandstreet2009}). The
stellar field truncates the inner disc, creating a low-density
magnetospheric cavity, where the matter ram pressures in the disc
balances the magnetic pressure of the stellar field (e.g.
\citealt{Konigl1991}). This density transition between the
low-density magnetosphere and the higher-density accretion disc
has a sharp positive density gradient due to the truncation by the
magnetic field (e.g. \citealt{RomanovaEtAl2002,RomanovaEtAl2003,
RomanovaLovelace2006}.

A migrating planet which encounters this boundary becomes trapped
due to this sharp density gradient at the disc-magnetosphere
interface. For a typical T Tauri star, the magnetospheric boundary
is located at $\sim 0.05-0.1$ AU and may serve as the final
barrier, preventing planets from migrating all the way into their
host stars. The accretion rates in the disc around young stars may
span several orders of magnitude, from $\mdot \sim 10^{-7} -
10^{-11} \msun/{\rm yr}$, meaning that the size of the
magnetospheric cavity can also differ substantially between young
stellar systems (e.g. \citealt{Hartmann2000}).

 If the disc
accretion rate declines and the magnetospheric radius increases,
the trapped planet will move outward with the inner edge of the
disc due to the action of the corotation torque
\citep{MassetEtAl2006a,LiuEtAl2017}.
 \citet{LiuEtAl2017} suggested that
this mechanism may be responsible for the dispersal of low-mass
planets at $r\lesssim 1$ AU and their observed nearly homogeneous
distribution with periods.

\subsection{A planet trap at the dust sublimation radius and at the dead zone}

Observations of protoplanetary discs show an infrared excess at 3
$\mu$m, which can be attributed to an inner disc edge at a
distance coinciding with the dust-sublimation radius of the system
\citep{DullemondEtAl2001}. Exterior to this radius, the disc is
cold and dusty, and the MRI-powered accretion (which is sensitive
to the ionization fraction in the disc) proceeds slowly.

However, interior to the dust sublimation radius, the gas disc is
optically thin and the ionization fraction is high. As a result,
the viscosity provided by the MRI turbulence is large and
accretion interior to the dust sublimation radius proceeds
efficiently. The result is an evacuated low-density inner cavity
surrounded by a cold, higher density disc. Migrating planets which
encounter this inner cavity may become trapped at the density
transition and migrate in or out if the dust sublimation radius
changes as the young star moves along the Hayashi track towards
the main sequence.

Infrared observations of protoplanetary discs show that these
inner cavities can range from $\sim$0.1 to 4 \AU\ in size,
depending on the luminosity of the star \citep{MuzerolleEtAl2003,
AkesonEtAl2005, MonnierEtAl2005}. The outer edge of this cavity
may act as a trap for inwardly migrating planetesimals, allowing
for the gradual buildup of planetary embryos in this region
\citep{MorbidelliEtAl2008}.

The low-mass migrating planets may also become trapped at the
inner edge of a dead zone
\citep{Gammie1996,MatsumuraEtAl2007,GuileraSandor2017}, where the
ionizing radiation does not penetrate far enough into the disc to
sustain MRI-driven accretion \citep{MassetEtAl2006a}. The surface
density in the dead zone is expected to be substantially higher
than the density in the inner regions, where the MRI-driven
turbulence is ongoing. As such, the inner edge of the dead zone
may serve as a trap for migrating planets. Like the dust
sublimation radius, the inner dead zone radius may change in time,
causing planets trapped at the inner edge to move in or out
correspondingly \citep{MassetEtAl2006a,KretkeLin2012}.

\subsection{Planet traps at snow lines}

The positive density gradients can also form at snow lines,
associated with the phase transition from liquid to frozen states.
Each molecular species has its own condensation front, depending
on its unique freeze-out temperature, and therefore several snow
lines are expected in protoplanetary discs.  At the snow line, a
sudden change in viscosity is expected due to a sharp increase in
the grain-to-dust density ratio, and the changes of the turbulence
properties (driven by MRI).

The snow lines associated with freezing of water are the closest
to stars and are expected to be at distance of several AU (e.g.
\citealt{KretkeLin2007}). The positive density gradient and the
corresponding corotation torque can stop Type I migration of
several Earth mass cores \citep{ZhangEtAl2007}, which are
necessary for the formation of giant planets
\citep{PollackEtAl1996} and their survival at 2-3 AU.

Recently, the $\sim 30$ AU snow line associated with the freezing
of CO molecules has been resolved in T Tauri star TW Hya by
\textit{ALMA} telescope (e.g. \citealt{QiEtAl2013}), which was a
proof that snow lines are a prominent features in protoplanetary
discs.

Note that at the snow lines,  and other types of disc-cavity
boundaries, the positive corotation torque can also be connected
with the negative entropy gradient (e.g.
\citealt{PaardekooperMellema2006,KleyCrida2008,KleyEtAl2009,PaardekooperPapaloizou2009c,HasegawaPudritz2011,LegaEtAl2014}).
This is an additional physical mechanism which may stop migration
of planets at the disc-magnetosphere boundary.


\section{Conclusions} \label{sec:conclusion}

We present global 3D simulations of low-mass planets migrating
near the edge of a disc cavity. The main points of this study are
as follows:
\begin{itemize}
  \item For planets migrating far from the cavity boundary, the migration proceeds due to the excitation of density waves
    at the Lindblad resonances. However, migrating planets which encounter the boundary experience a large positive
    corotation torque due the the positive surface density gradient at the disc-cavity interface.
    \item There exists a region in the disc where the magnitude of the corotation torque is equal and
    opposite to the differential Lindblad torque, resulting in a region where a planet experiences zero
    net torque from the disc. Interior to this region, the net torque is positive due to the corotation
    torque and the planet migrates outward; similarly, exterior to this location, differential Lindblad
    torque is larger in magnitude and the net torque is negative, resulting in inward migration.
     Hence the zero-torque region is a stable ``planet trap'' as any perturbations in either direction
     result in a torque which pushes the planet back toward the trapping region.
    \item An accretion disc around a young star may have several planet traps wherever
    there is a large positive surface density gradient in the disc. Such density
    gradients
    may appear   at the edge
    of the magnetospheric cavity,     at the inner edge of the dead zone, at the dust-sublimation radius, or at snow lines.
    As the disc's accretion rate or the stellar luminosity change in time,
    the locations of these boundaries may shift. Planets which are trapped at the density gradients
    are likely to migrate along with the boundary.
    \item The observed dispersed distribution of low-mass planets at $r\lesssim 1$ AU may be connected with trapping of migrating planets
     at different distances from their host stars.
\end{itemize}

\section*{Acknowledgments}
Resources supporting this work were provided by the NASA High-End
Computing (HEC) Program through the NASA Advanced Supercomputing
(NAS) Division at Ames Research Center and the NASA Center for
Computational Sciences (NCCS) at Goddard Space Flight Center. The
research was supported by  NASA grant NNX12AI85G. AVK and GVU were
supported by the RFBR grant 18-02-00907.

\bibliographystyle{mn2e}

\appendix

\section{Dependence on the disc aspect ratio}
\label{sec:appen-aspect ratio}

\begin{figure}
    \centering
        \includegraphics[width=0.45\textwidth]{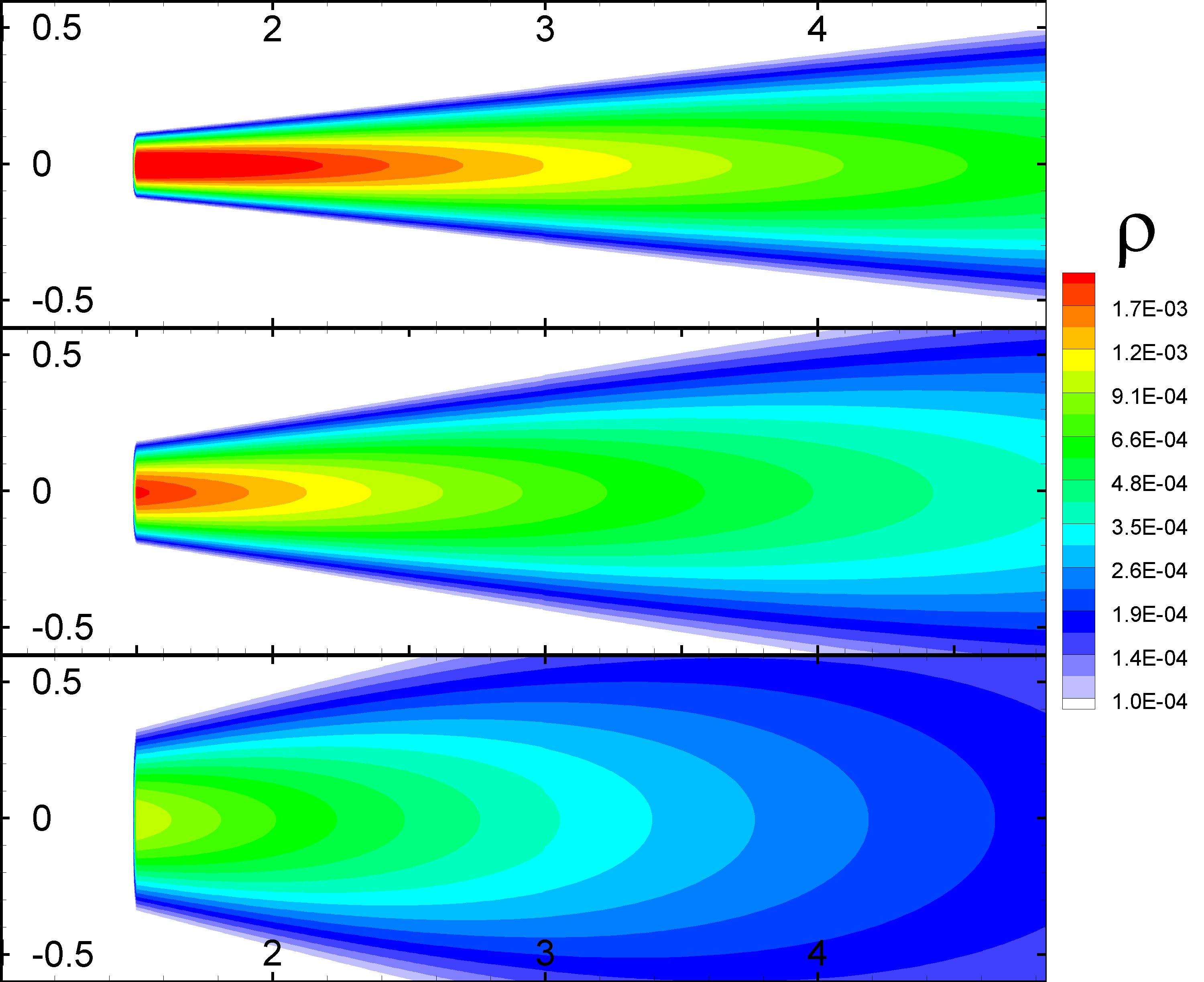}
    \caption{Discs with different aspect ratios. From top to bottom: density distribution in the $rz-$plane in three discs with differing aspect ratios of $H/r$ = 0.03, 0.05, and 0.10 but
     identical surface density profiles. \label{fig:asp_ratio_lowmass}}
\end{figure}

\begin{figure*}
    \centering
        \includegraphics[width=.9\textwidth]{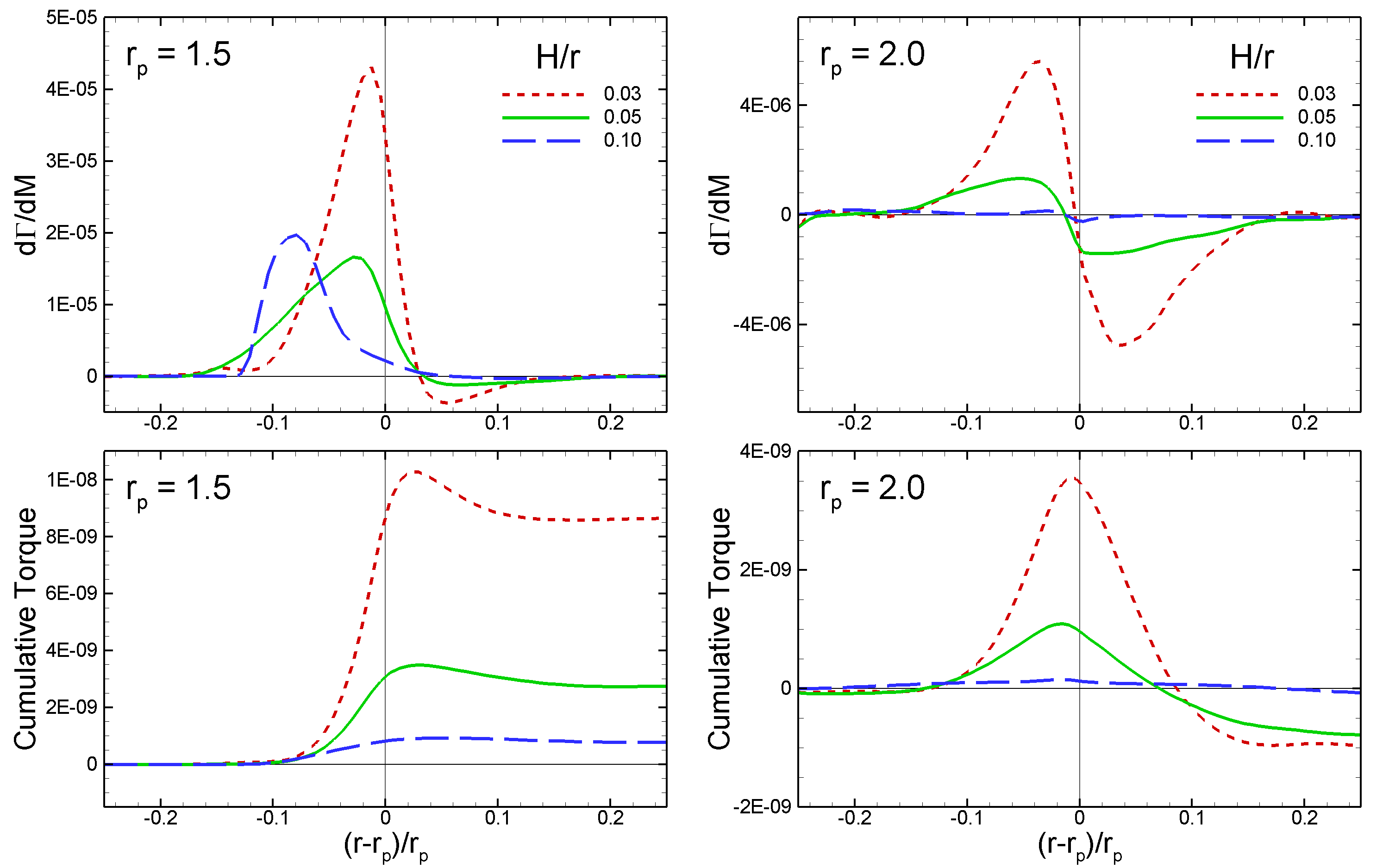}
    \caption{Torque profiles for different disc aspect ratios for a planet with mass $M_p=15 M_\oplus$.
    \textit{Top row:} the torque per unit mass as a function
    of normalized radius for planets starting at $r_p$ = 1.5 and 2.0 in three discs with differing aspect ratios, but identical surface density profiles. \textit{Bottom row:}
    same as the top row, but showing the cumulative torque as a function of normalized radius. The profiles are time averaged over $t = 70$.
    \label{fig:dTdM_asp_ratio}} 
\end{figure*}

In this study, we chose a relatively small initial aspect ratio,
$h/r=0.03$, which corresponds to  cold thin discs expected in
protoplanetary systems. However, in some instances, the inner
parts of the disc can be thicker due to stellar irradiation, or
some other reasons. In order to investigate the effect of the disc
height on planet migration, we performed calculations  at larger
initial disc aspect ratios: $H/r=0.05$ and $0.1$. For that, we
fixed the disc's surface density profile at $\Sigma=4 \times
10^{-4}=const$ and also fixed the density (and pressure)
distributions in the equatorial plane as $\rho_{eq}\sim r^{-3/2}$,
$p_{eq}\sim r^{-3/2}$. Then, the fiducial densities in the discs
with aspect ratios $0.05$ and $0.1$ are $\rho_{\rm d}=2.0\times
10^{-3}$ and  $1.0\times10^{-3}$, respectively. The fiducial
density in the cavity is taken to be $\rho_{\rm cav}=10^{-2}
\rho_{\rm d}$.
\fig{fig:asp_ratio_lowmass} shows the resulting density
distribution for discs with  $H/r$ = 0.03, 0.05, and 0.1.

In models with larger aspect ratios, we increased the size of the
simulation region in $z-$direction: in the cases of the discs with
an aspect ratio of $H/r=0.05$, the region spans \tz = -1.0 to
+1.0, and the number of grid cells $N_z=160$. In the cases of the
discs with an aspect ratio of $H/r=0.1$, the region spans \tz =
-2.0 to +2.0, and the number of grid cells $N_z=320$.

To compare the torques acting on the planets in discs with
different $H/r$, we measure the torques on the planets located at
$r_p = 1.5$ (the disc-cavity boundary) and $r_p = 2.0$ (in the
disc). \fig{fig:dTdM_asp_ratio} shows the torque per unit mass
(top row) and cumulative torque (bottom row) on the  planet as a
function of normalized radius, time averaged from $t = 10 - 80$.
Evidently, the net torque on the planet is diminished by
increasing the disc's aspect ratio for both the planet at $r_p =
1.5$ where the corotation torque is important and at $r_p = 2.0$
where the differential Lindblad torque dominates. The bottom
panels of \fig{fig:dTdM_asp_ratio} show that the tidal torque's
dependence on the aspect ratio is nonlinear, with much stronger
torques in thinner discs. This is due to the fact that as the disc
thickness increases, less mass is concentrated near the disc
midplane. For the corotation resonance, this means that less
matter participates in the horseshoe orbit and the magnitude of
the corotation torque is smaller. Similarly, for the Lindblad
torques, the amplitude of the surface density perturbations raised
by the planet is smaller and the torque again weakens.

The implication of these results is that for reasonable values of
$H/r$, varying the disc height does not change the direction of
the migration; however, it may strongly modify the migration time
scale of the planet. This effect is especially significant for the
planets interacting with the disc-cavity boundary where the
corotation torque is important. However, since the migration
direction is unaffected, the planet trapping mechanism is robust
across the varying disc aspect ratios.

\section{Dependence on viscosity}
\label{sec:appen-viscosity}

In viscous discs, the disc matter moves inward, enters the
horseshoe region and can prevent the corotation torque from
saturation. To prevent saturation, the viscous time scale across
the horseshoe region should be smaller than  the horseshoe
libration time scale (e.g. \citealt{Masset2001}):
\begin{equation}
\frac{x_s^2}{3\nu_{\rm vis}} < \frac{4\pi r_p}{(3/2)\Omega_p x_s}
~.
\end{equation}
Following \citet{Masset2001}, we take the value of the half-width
horseshoe region in the form of $x_s=0.96 r_p\sqrt{{q_p}/{h}}$ and
obtain the minimum value of viscosity required for saturation:
\begin{equation}
\nu_{\rm vis}=0.035\bigg(\frac{q_p}{h}\bigg)^{3/2} r_p^2\Omega_p
~.
\end{equation}
Taking into account that $\nu_{\rm vis}=\alpha H^2\Omega_p$, we
obtain the minimum value of the $\alpha-$parameter of viscosity:
\begin{equation}
\alpha=0.035 q^{3/2} h^{-7/2}\approx 4.3\times 10^{-4} \bigg(
\frac{M_p}{5 M_\oplus}\bigg)^{3/2} \bigg(
\frac{h}{0.03}\bigg)^{-7/2}~.
\end{equation}
For planets with masses  $5 M_\oplus$ and $15 M_\oplus$, the
minimum values of $\alpha$ required for the unsaturation of the
corotation torque are $\alpha\approx 4.5\times10^{-4}$ and
$\alpha\approx 2.2\times 10^{-3}$, respectively.

\begin{figure}
\centering
\includegraphics[width=0.5\textwidth]{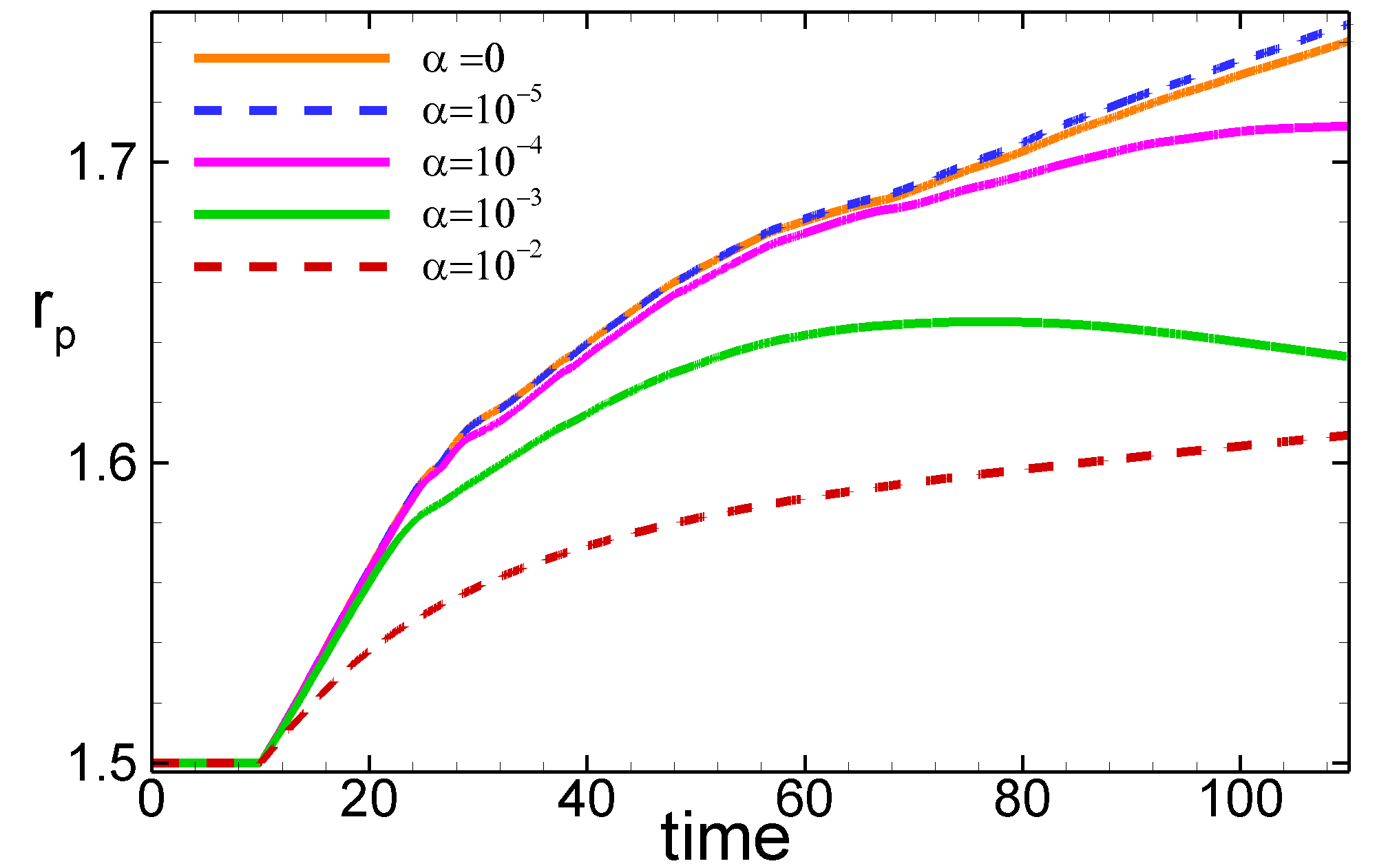}
\caption{Variation of the planet's semi-major axis $r_p$ in time
in simulations with different values of viscosity parameter
$\alpha$ in the disc for a planet with mass $M_p=15 M_\oplus$ and
initial value $r_p=1.5$. } \label{fig:rp-alpha}
\end{figure}

\begin{figure*}
    \centering
        \includegraphics[width=0.33\textwidth]{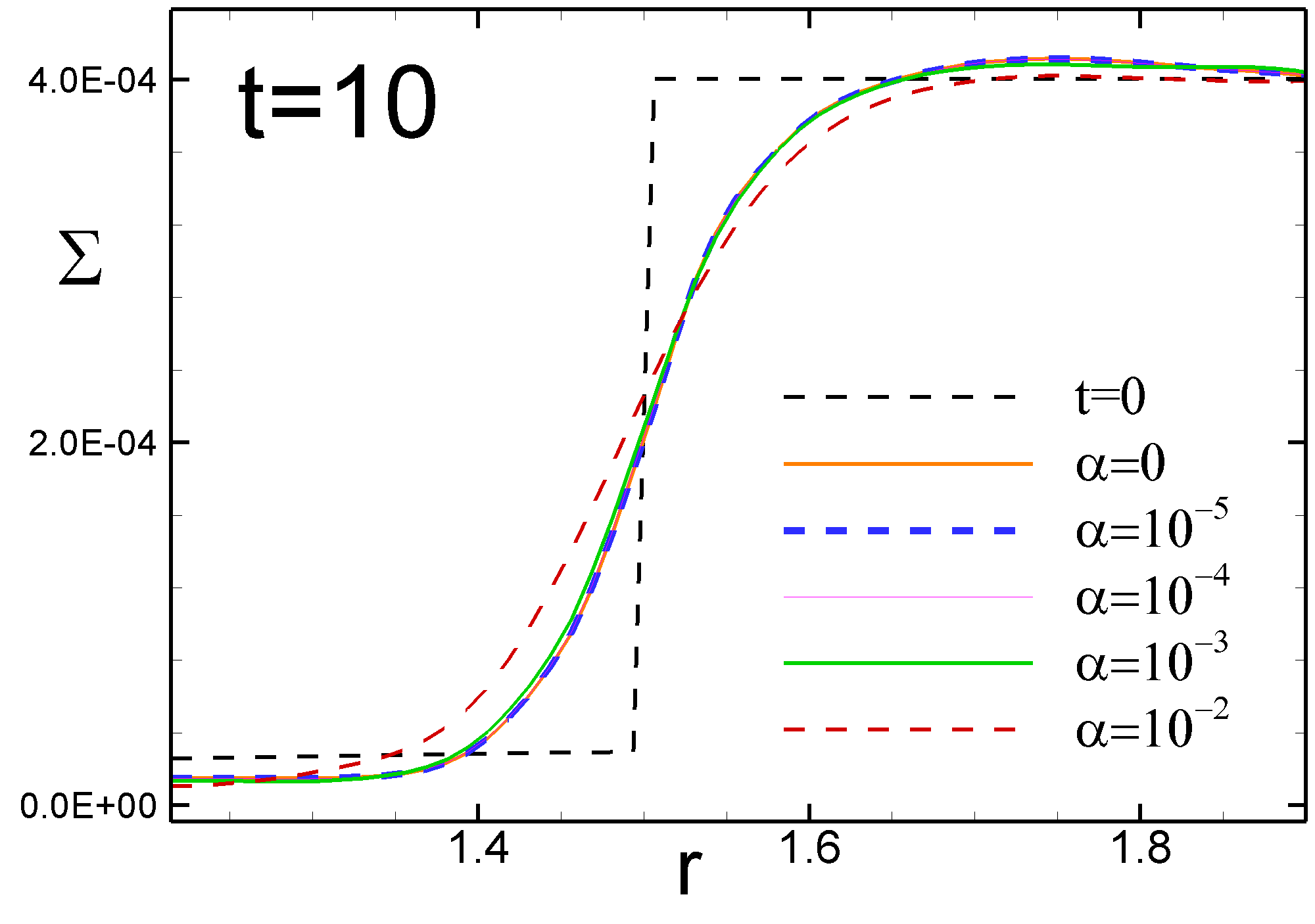}
          \includegraphics[width=0.33\textwidth]{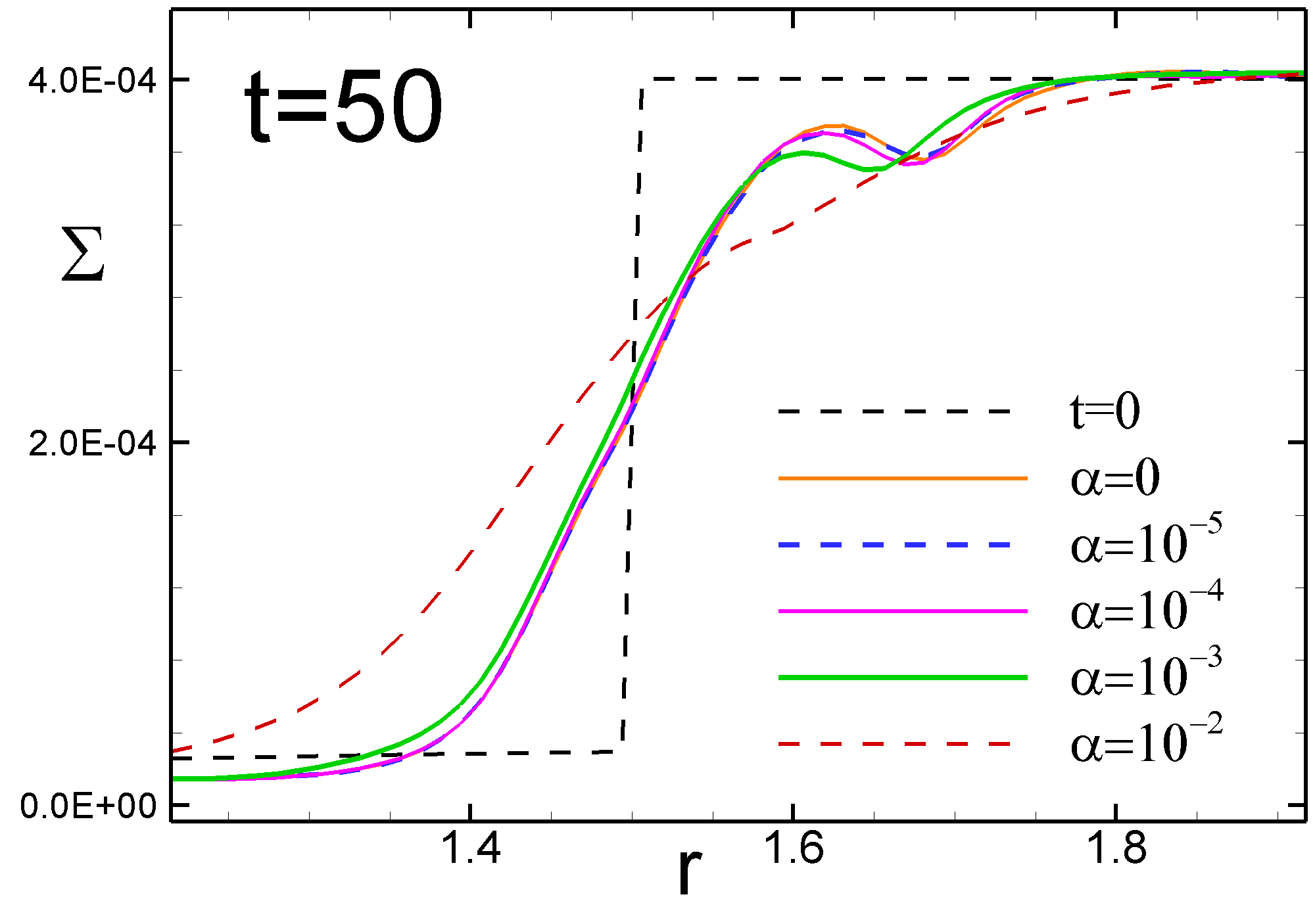}
        \includegraphics[width=0.33\textwidth]{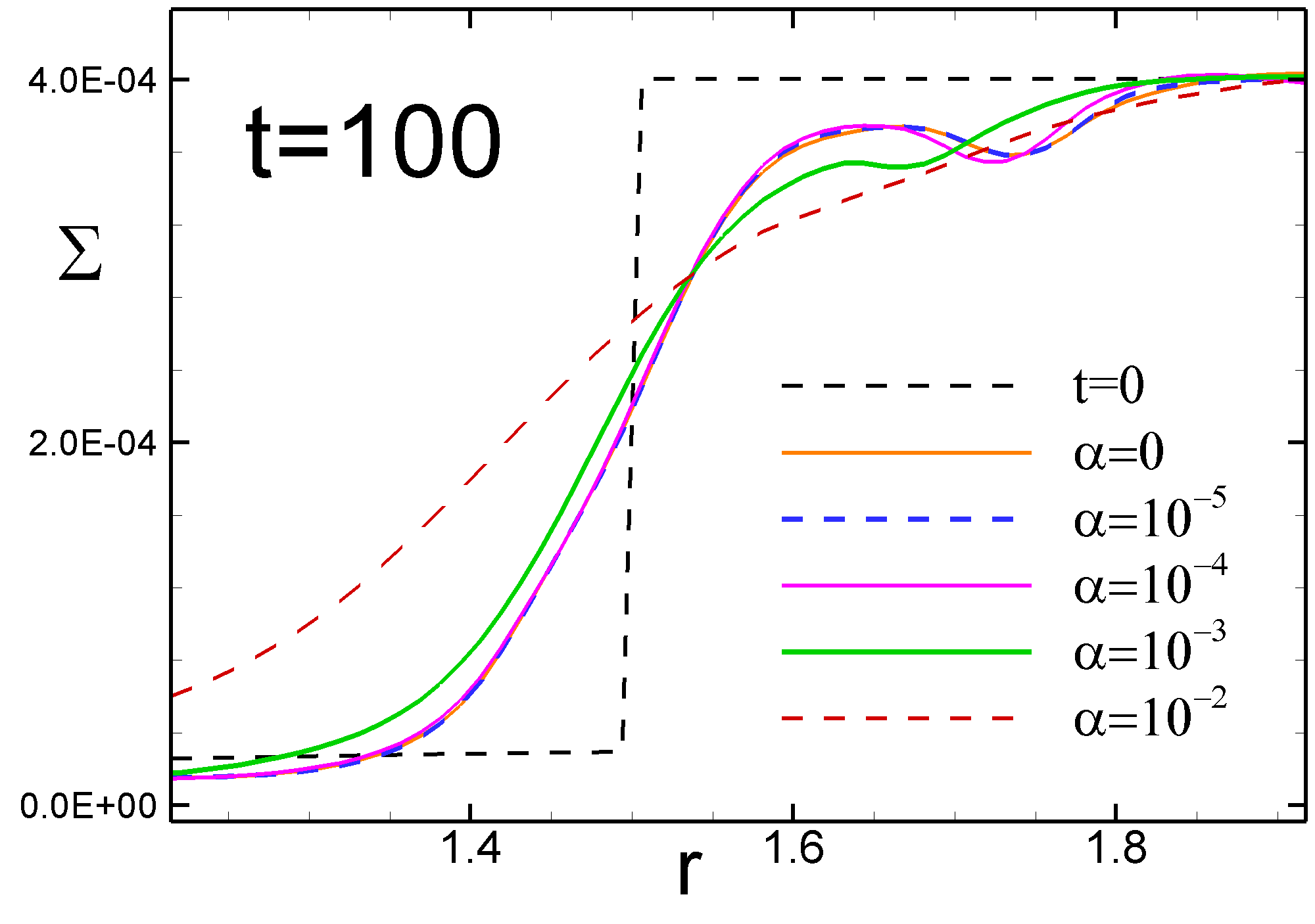}
    \caption{The surface density distribution in the vicinity of the disc-cavity boundary
    for three moments in time, $t=10, 50, 100$, for different
    parameters of $\alpha-$viscosity in the disc and a planet with mass $M_p=15 M_\oplus$.
     }. \label{fig:sigma-alpha-3}
\end{figure*}

Our main simulations were performed at zero viscosity. Here, we
test the role of viscosity in our model. We switched on the module
of viscosity and investigated the migration of a planet with mass
$15M_\oplus$, located  at $r_p=1.5$, in discs with different
values of the viscosity parameter: $\alpha=10^{-5}, 10^{-4},
10^{-3}$ and $10^{-2}$. We observed that, in all cases, the planet
migrates outward (see Fig. \ref{fig:rp-alpha}). However, the
migration rate is highest at the lowest values of the viscosity
parameter $\alpha$. This dependence is the opposite of that
expected in the cases where the viscosity is responsible for the
unsaturation of the corotation torque. The reason for this
dichotomy is specific to our model: the corotation torque is
unsaturated due to the rapid migration of the planet from the
disc-cavity boundary, while accretion due to viscosity is a slower
process. In addition, in the presence of viscosity, the
disc-cavity boundary becomes wider, and therefore the density
gradient (and the migration rate) becomes lower. Fig.
\ref{fig:sigma-alpha-3} shows that the density gradient does not
change significantly if $\alpha \lesssim 10^{-3}$. Therefore, our
quasi-steady model of the disc-cavity boundary is only applicable
in the cases of low values of $\alpha$.

In the above example, a planet migrates rapidly from the region of
the  high density gradient. If a planet is located in the part of
the disc with a lower density gradient, then the outward migration
rate is lower, and the viscosity may dominate in the unsaturation
of the corotation torque.

\end{document}